\documentclass[twocolumn,amsthm]{autart}    

\usepackage{graphicx}          
\usepackage{amsmath}
\usepackage{graphicx}
\usepackage[table,xcdraw]{xcolor}

\usepackage{pgfplots}
\pgfplotsset{compat=1.17}

\usepackage[switch]{lineno}
\usepackage{multirow}
\usepackage{amssymb}
\usepackage{dsfont}
\usepackage{enumerate}
\usepackage{xcolor}
\definecolor{mycolor}{rgb}{0, 1, 0}
\definecolor{mycolor2}{rgb}{0, 0, 0}
\newcommand{\playerone}{1}
\newcommand{\playertwo}{2}
\newcommand{\average}{\mathsf{Av}}
\newcommand{\states}{\mathcal{S}}
\newcommand{\cardstates}{S}
\newcommand{\cardactions}{A}
\newcommand{\winningstates}{\states^{R}}
\newcommand{\actions}{\mathcal{A}}
\newcommand{\probs}{\mathcal{P}}

\newcommand{\initialstate}{s_{0}}
\newcommand{\histories}{\mathcal{H}}
\newcommand{\history}{h}
\newcommand{\policy}{\pi}
\newcommand{\Policy}{\Pi}

\newcommand{\discount}{\beta}
\newcommand{\policies}{\Pi}

\newcommand{\genericset}{D}
\newcommand{\samplepaths}{n}
\newcommand{\sampleperstate}{m}
\newcommand{\expectation}{\mathbb{E}}
\newcommand{\simplex}{\Delta}
\newcommand{\genericstate}{s}
\newcommand{\altstate}{q}
\newcommand{\genericaction}{a}
\newcommand{\maxcost}{c_{max}}
\newcommand{\known}{K}
\newcommand{\knownstates}{\states^{\known}}
\newcommand{\unknownstates}{\states^{U}}
\newcommand{\trapstates}{\states^{trap}}
\newcommand{\statesend}{\states^{end}}
\newcommand{\potentiallywinningstates}{\states^{+}}
\newcommand{\failureprob}{\delta}
\newcommand{\identityconcealmentgame}{\mathcal{IC}}
\newcommand{\optimalitygap}{\varepsilon}
\newcommand{\losingprobability}{\lambda}
\newcommand{\gamevalue}{v}

\newcommand{\payoff}{c}
\newcommand{\randomstoppingtime}{\tau}
\newcommand{\absorbingstates}{\states^{abs}}

\usepackage{algorithm}
\usepackage{algpseudocode}
\usepackage{tikz}
\usepackage{booktabs}
\usepackage{bbold}
\usetikzlibrary{automata, positioning, arrows}
\usepackage[utf8]{inputenc}
\usepackage[english]{babel}
\usepackage{mathtools}
\usepackage{natbib}

\usepackage{tikz}
\usetikzlibrary{automata, positioning, arrows}

\definecolor{purple}{rgb}{1,0.5,1}
\definecolor{green}{rgb}{0.5,1,0.5}
\definecolor{red}{rgb}{1,0.5,0.5}

\usepackage{subcaption}

\graphicspath{{images/}}

\algtext*{EndWhile}
\algtext*{EndFor}
\algtext*{EndFunction}
\algtext*{EndIf}

\allowdisplaybreaks

\usepackage{xpatch}

\makeatletter 
\xpatchcmd{\runningauthor@fmt}{\global\edef}{\protected@xdef}{}{}
\xpatchcmd{\runningauthor@fmt}{\global\edef}{\protected@xdef}{}{}
\xpatchcmd{\author@fmt}{\edef}{\protected@edef}{}{}
\def\@xnamedef#1{\expandafter\protected@xdef\csname #1\endcsname}
\def\ead@au#1{\protected@edef\@ead@au{#1}}
\def\add@xtok#1#2{\begingroup
  \protected@xdef\@act{\global\noexpand#1{\the#1#2}}\@act
\endgroup}
\def\no@harm{}
\makeatother

\begin{document}

\begin{frontmatter}

\title{Identity Concealment Games: How I Learned to Stop Revealing and Love the Coincidences\thanksref{footnoteinfo}} 

\thanks[footnoteinfo]{This paper was not presented at any IFAC 
meeting. Corresponding author M.~O.~Karabag.}

\author[UTECE]{Mustafa O. Karabag}\ead{karabag@utexas.edu},    
\author[UIUCASE]{Melkior Ornik}\ead{mornik@illinois.edu},               
\author[UTASE]{Ufuk Topcu}\ead{utopcu@utexas.edu}  

\address[UTECE]{Oden Institute for Computational Engineering and Sciences, The University of Texas at Austin, TX, USA}  
\address[UIUCASE]{Department of Aerospace Engineering, University of Illinois at Urbana–Champaign, IL, USA}             
\address[UTASE]{Department of Aerospace Engineering and Engineering Mechanics, The University of Texas at Austin, TX, USA}        


\begin{abstract}
In an adversarial environment, a hostile player performing a task may behave like a non-hostile one in order not to reveal its identity to an opponent. To model such a scenario, we define identity concealment games: zero-sum stochastic reachability games with a zero-sum objective of identity concealment. To measure the identity concealment of the player, we introduce the notion of an average player. The average player's policy represents the expected behavior of a non-hostile player. We show that there exists an equilibrium policy pair for every identity concealment game and give the optimality equations to synthesize an equilibrium policy pair. If the player's opponent follows a non-equilibrium policy, the player can hide its identity better. For this reason, we study how the hostile player may learn the opponent's policy. Since learning via exploration policies would quickly reveal the hostile player's identity to the opponent, we consider the problem of learning a near-optimal policy for the hostile player using the game runs collected under the average player's policy. Consequently, we propose an algorithm that provably learns a near-optimal policy and give an upper bound on the number of sample runs to be collected.
\end{abstract}

\end{frontmatter}
\section{Introduction}

In an adversarial environment, an agent interacts with a non-cooperative opponent. For a hostile agent, it may be important not to expose its identity since the opponent might attempt to hinder the agent's operation knowing that the agent is hostile. For instance, intelligence services often instruct the agents who are under surveillance to \textit{dry-clean}, that is, to evade surveillance in a way that looks accidental, not intentional, since intentional evasions cause suspicion~\citep{macintyre2019}. This behavior motivated video games such as Spy Party~\citep{spyparty} and Garry's Mod Guess Who~\citep{guesswho} where the goal is to complete tasks behaving like a non-playable character, i.e., a bot. While identity concealment is a significant behavior in reality, it has not been studied in the literature of zero-sum games, which is a common formalism of adversarial settings~\citep{kardes2005survey}.

\begin{figure}[t]
    \centering
    \includegraphics[scale=0.35]{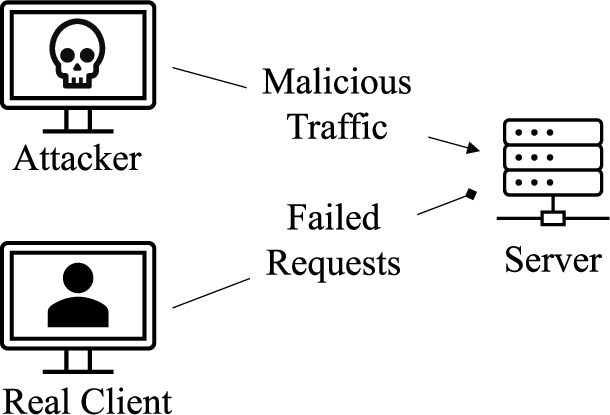}
    \caption{A cyber interaction scenario as a two-player game. The game is played between a client and the server. The server does not know the identity of the client, whether the client is an attacker or a real client. The states of the game represent the remaining times for the client's processed requests if there are any. At every time, the client can disconnect, make a request or wait. The server can accept or reject the client's potential request.}
    \label{fig:dosattack}
\end{figure}

We formalize the above notion of \textit{identity concealment} in two-player zero-sum reachability games. We consider a graph as the state space of the game. The goal of a \textit{hostile player} is to reach a set of target states, but in a way that its behavior looks similar to the behavior of non-hostile players, i.e., the hostile player aims to make its win look coincidental. The goal of the opponent is to distinguish between hostile and non-hostile players. As a reference point, we introduce an abstract notion of an \textit{average player} to measure the identity concealment of a hostile player. The average player's policy represents the expected behavior of non-hostile players. For example, in the cyber interaction scenario shown in Figure \ref{fig:dosattack}, hostile players are attackers who perform a denial-of-service attack against a server, and average players are real clients interacting with the server. The attackers' goal is to overwhelm the server and make it fail to provide service to real clients while not being identified. The server is the opponent that aims to distinguish the attackers from real clients. We measure identity concealment by the cumulative Kullback-Leibler (KL) divergence between the action distribution of a hostile player and the action distribution of an average player over a game run. As the KL objective function increases, the opponent can distinguish a hostile player from a non-hostile player more easily. For example, in the cyber interaction scenario, a game run can represent the history of the client's requests and the server's responses. In particular, possibly complex or time-varying behavior (such as repeated requests for unrelated computationally heavy resources) may indicate that the client is hostile, i.e., an attacker.

We define the \textit{identity concealment game} as the two-player zero-sum reachability game with the cumulative KL divergence objective function. We first identify the conditions for which the value of the KL objective function is finite, i.e., the opponent is never sure the player is hostile. Then, we show that there exists an equilibrium policy pair for a hostile player and the opponent, which can be synthesized using value iteration. 

The hostile player can achieve a lower value than the equilibrium value if the opponent follows a non-equilibrium policy. In this case, an equilibrium policy is not necessarily optimal for the hostile player against an imperfect opponent. The hostile player needs to learn and respond to the opponent's suboptimal policy to achieve the optimal value. However, the player's ability to learn in the described setting is limited in that an active learner would quickly reveal its identity during exploration. We consider the question of whether it is possible to learn a near-optimal policy offline by solely using the game runs collected under the average player's policy. The output policy needs to be near-optimal in that the KL objective function is \(\varepsilon\)-optimal, and the probability of winning is at least \(1-\lambda\) where \(\varepsilon\) and \(\lambda\) are the input parameters of the algorithm.

We provide an algorithm that solely uses a finite number of runs collected under the average player's policy to learn a near-optimal policy. To show the near-optimality in the KL objective, we utilize and improve some of the probably approximately correct Markov decision processes (PAC-MDP) learning results~\citep{fiechter1994efficient,kearns2002near,strehl2008analysis}. To show the near-optimality in the probability of winning, we show that under the output policy, the hostile player can lose the game only if an unknown state, i.e., a state with a low number of samples, is visited. Then, we show that the unknown states cannot be visited with a high probability if the number of sample runs is high enough.

We give the proofs of some technical results in \citep{karabag2021identity} due to lack of space.

\section{Related Work}
The KL objective function is used for different purposes including \textit{deception in supervisory control} \citep{karabag2021deception}, \textit{game balancing} \citep{grau2018balancing}, \textit{inverse reinforcement learning}~\citep{boularias2011relative}, and \textit{reinforcement learning}~\citep{fox2016taming,peters2010relative}. \cite{karabag2021deception} utilized Sanov's theorem~\citep{cover2012elements} and the KL divergence of the path distributions in MDPs for deception in supervisory control. In that paper, the supervisor designs a reference policy to an agent, which is supposed to follow this policy, but it deviates from the reference policy to achieve a malicious task. The goal of the supervisor is to design a reference policy that minimizes deviations. While we use the objective function for the same purpose, this paper differs from \citep{karabag2021deception} in that the opponent (analogue of the supervisor) does not design the average player's policy (analogue of the reference policy). Instead, the opponent designs a policy that determines the observability of the player (analogue of the agent). \cite{grau2018balancing} used the KL divergence objective for game balancing in two-player stochastic games. Aside from the contextual differences, the objective function in  \citep{grau2018balancing} has a discount factor. We, on the other hand, do not have a discount factor that significantly differs the proof for the existence of an equilibrium. Goal and plan obfuscation~\citep{kulkarni2019unified,keren2016privacy} are similar to the concept of identity concealment. We consider a measure based on statistical hypothesis testing, whereas the cited works consider measures based on the distance of the observation sequences generated by a game run.

The learning algorithm provided in this paper is related to PAC-MDP algorithms~\citep{kearns2002near,strehl2008analysis}. While these algorithms guarantee near-optimality after a finite number of suboptimal actions, there are no guarantees on the suboptimality of the transient learning period due to exploration. In the adversarial setting described in this paper, the use of PAC-MDP algorithms would reveal the identity of the player during the learning period. The algorithm provided in this paper uses a fixed policy, the average player's policy, to learn, whereas PAC-MDP algorithms learn in an exploratory manner. The learning algorithm provided in this paper is also related to off-policy evaluation and optimization~\citep{farajtabar2018more,precup2000eligibility,yu2020mopo,kidambi2020morel,levine2020offline} as we collect offline samples using a behavior policy that is not the target policy to be evaluated or optimized. In detail, our algorithm is similar to model-based off-policy optimization~\citep{yu2020mopo,kidambi2020morel}. The existing offline learning literature considers the single-objective discounted infinite horizon~\citep{puterman2014markov} setting. On the other hand, we consider a multi-objective infinite horizon setting, where one of the objectives, probability of winning, is undiscounted, and the other objective is the KL divergence. Similar to the existing model-based offline learning methods, we show the near-optimality in the KL divergence by showing that the learned model is close to the actual model. To show the near-optimality in the probability of winning, we use a new approach that utilizes the near-optimality of the KL objective function.

The sample complexity of offline policy optimization is dependent on the distributional shift between the behavior and optimal policies~\citep{levine2020offline}. However, quantifying the distributional shift is challenging since it requires knowing the statistical properties of the processes induced by the policies. To ensure that the learned policy does not suffer from distributional shift, existing learning methods use KL divergence as a regularizer~\citep{schulman2015trust}. We, on the other hand, use the equilibrium value of KL divergence to reason about the maximum distributional shift that a near-optimal policy can have and guarantee near-optimality in the probability of winning: Since the maximum distributional shift of the near-optimal policy is bounded, we can bound the probability of losing for the hostile player. This approach allows us to give an explicit bound on the number of samples required to synthesize a near-optimal policy. Unlike the existing model-based off-policy optimization works that provide sample complexity bounds with agnostic dependencies on the distributional shift~\citep{ross2012agnostic,uehara2021pessimistic}, we give a bound that has a known dependency on the distributional shift thanks to the known equilibrium value of the game.

	\section{Preliminaries} \label{section:preliminaries}
In this section, we give preliminaries on two-player stochastic games and objective functions for the games.

\subsection{Two-Player Stochastic Games and Markov Decision Processes}

A \textit{two-player stochastic game} is a tuple \( \mathcal{G} = (\states, \actions^{\playerone}, \actions^{\playertwo}, \allowbreak \probs,  \initialstate)\) where \(\states\) is a finite set of states, \(\actions^{\playerone}\) is a finite set of actions for Player 1, \(\actions^{\playertwo}\) is a finite set of actions for Player 2, \(\probs:\states \times \actions^{\playerone} \times \actions^{\playertwo} \times \states \to [0,1]\) is the transition probability function, and \(\initialstate \in \states\) is the initial state. We note that the available actions can be state-dependent. We use fixed sets, \(\actions^{\playerone}\) and \(\actions^{\playertwo}\), for the available actions of players for clarity of presentation. For every state \(\genericstate \in \states\),  \(\sum_{q \in \states} \probs(\genericstate,a^{\playerone}, a^{\playertwo}, q) = 1\) for all \((a^{\playerone} ,a^{\playertwo}) \in \actions^{\playerone} \times \actions^{\playertwo}\). We use \(\cardstates\) to denote the cardinality of \(\states\), and \(\cardactions\) to denote the maximum of the cardinalities of \(\actions^{\playerone}\) and \(\actions^{\playertwo}\). The \textit{successor states} of state \(\genericstate\) is denoted by \(Succ(\genericstate)\) where state \(q\) is in \(Succ(\genericstate)\) if and only if there exist actions \((a^{\playerone} ,a^{\playertwo}) \in \actions^{\playerone} \times \actions^{\playertwo}\) such that \(\probs(\genericstate,a^{\playerone}, a^{\playertwo}, q) > 0\). \textcolor{mycolor2}{State \(\genericstate\) is \textit{absorbing} if and only if \(Succ(\genericstate) = \lbrace \genericstate \rbrace\), and there is only one available action for each player. The set of all absorbing states is  \(\absorbingstates\).
}

The game has infinite steps. At every time step \(t\), Players \(1\) and \(2\) choose their actions, \(a^{\playerone}_{t}\) and \(a^{\playertwo}_{t}\), simultaneously and transition to state \(s_{t+1}\) from \(s_{t}\) with probability \(\probs(s_{t},a^{\playerone}_{t}, a^{\playertwo}_{t}, s_{t+1})\). The history \(h_{t} = s_{0}a^{\playerone}_{0}a^{\playertwo}_{0}\ldots s_{t-1}a^{\playerone}_{t-1} a^{\playertwo}_{t-1} s_{t}\) at time \(t\) is the sequence of all previous states and actions. The set of all possible histories at time \(t\) is denoted by \(\histories_{t}\). 

A \textit{(history-dependent) policy }for Player \(i\) is a sequence \(\policy^{i} = \mu^{i}_0\mu^{i}_1\ldots\) where each \(\mu^{i}_t:\histories_{t} \times \actions^{i} \to [0,1]\) is a decision function such that \(\sum_{a^{i} \in \actions^{i}} \mu^{i}_t(\history_{t},a^{i}) = 1\)  for all \(\history_{t} \in \histories_{t}\). Given the history \(\history_{t},\) we use \(\mu^{i}_{t}(\history_{t})\) to denote the action distribution under Player \(i\)'s policy \(\policy^{i}\) at time \(t\) and state \(\genericstate_{t}\). A \textit{stationary policy} for Player \(i\) is a sequence \(\policy^{i} = \mu^{i} \mu^{i} \ldots\) such that \(\mu^{i}:\states \times \actions^{i} \to [0,1]\)  and \(\sum_{a^{i} \in \actions^{i}} \mu^{i}(\genericstate,a^{i}) = 1\) for all \(\genericstate \in \states\). The set of all policies for Player \(i\) is denoted by \(\policies^{i}\). The set of all stationary policies Player \(i\) is denoted by \(\policies^{i, St}\). For state \(\genericstate\), we use \(\policy^{i}(\genericstate)\) to denote the action distribution under Player \(i\)'s stationary policy \(\policy^{i}\). A \textit{run} \(\gamma = s_{0}a^{\playerone}_{0}a^{\playertwo}_{0}s_{1}a^{\playerone}_{1} a^{\playertwo}_{1}  \ldots\)  is an infinite sequence states and actions under policies \(\policy^{\playerone}\) and \(\policy^{\playertwo}\) such that \( \probs(s_{t},a^{\playerone},a^{\playertwo},s_{t+1})\) \( \mu^{\playerone}_t(h_{t},a^{\playerone}) \) \( \mu^{\playertwo}_t(h_{t}, a^{\playertwo}) > 0\) for all \(t\geq0\), i.e., all transitions are feasible. The probability distribution of runs under \(\policy^{\playerone}\) and \(\policy^{\playertwo}\) is denoted by \(\Gamma^{\policy^{\playerone}, \policy^{\playertwo}}\). The probability distribution of histories at time \(t\) is denoted by \(\Gamma^{\policy^{\playerone}, \policy^{\playertwo}}_{t}\). \textcolor{mycolor2}{The \textit{(undiscounted) occupancy measure} of state \(\genericstate\) is the expected number of times state \(\genericstate\) is visited and is equal to
\( \sum_{t=0}^{\infty} \Pr^{\policy^{\playerone}, \policy^{\playertwo}}(\genericstate_{t} = \genericstate|\genericstate_{0}) \).}

A \textit{Markov decision process} (MDP) is a tuple \(\mathcal{M}=(\states^{'}, \actions^{'}, \probs^{'}, \initialstate^{'})\) where \(\states^{'}\) is a finite set of states, \(\actions^{'}\) is a finite set of actions, \(\probs^{'}:\states^{'}\times \actions^{'} \times \states^{'} \to [0,1]\) is the transition probability function, and \(\initialstate^{'}\) is the initial state. A two-player stochastic game where one of the players uses a known, fixed policy is an MDP.

\subsubsection{Zero-Sum Objective and Equilibrium Policies}
A payoff function \(\payoff:\states\times\simplex_{|\actions^{1}|}\times\simplex_{|\actions^{2}|} \to \mathbb{R}\) maps the state and action distributions of the players to a payoff value where \(\simplex_{k}\) is the \(k\)-dimensional probability simplex. At time \(t\), Players \(\playerone\) and \(\playertwo\) with policies \(\policy^{1} = \mu^{1}_0\mu^{1}_1\ldots\) and \(\policy^{2} = \mu^{2}_0\mu^{2}_1\ldots\) receive a payoff of \(\payoff(\genericstate_{t}, \mu^{\playerone}_{t}(\history_{t}), \mu^{\playertwo}_{t}(\history_{t}))\). 

\textcolor{mycolor2}{Let \(X_{t} = (\genericstate_{t}, a^{\playerone}_{t},a^{\playertwo}_{t})\) be a random variable consisting of the random state and actions of the players at time \(t\). For the random process \(\lbrace X_{t} \rbrace\), the \textit{hitting time} \(\randomstoppingtime\) to the set \(\absorbingstates\) of absorbing states is a random variable defined by \(\randomstoppingtime = \min \lbrace t \geq 0 | s_{t} \in \absorbingstates\rbrace\) taking values in \(\mathbb{N} \cup \lbrace \infty\rbrace\). Using the hitting time, the zero-sum objective function \[C(\policy^{1}, \policy^{2}) = \expectation\left[ \sum_{t=0}^{\randomstoppingtime} \payoff(\genericstate_{t}, \mu^{\playerone}_{t}(\history_{t}), \mu^{\playertwo}_{t}(\history_{t})) \right]\] is the expected cumulative payoff until the random stopping time \(\randomstoppingtime\) where the expectation is over the randomness of policies, \(\policy^{1}\) and \(\policy^{2}\), and the dynamics \(\probs\) of the game. Player \(\playerone\) is the \textit{minimizer}, and Player \(\playertwo\) is the \textit{maximizer} of the zero-sum objective.}

Let \(P^{1}\) and \(P^{2}\) denote the fixed sets of feasible policies for Players \(\playerone\) and \(\playertwo\), respectively. A pair \((\policy^{1,*}, \policy^{2,*})\in P^{1} \times P^{2}\) of policies is an \textit{equilibrium policy pair} if and only if \[\sup_{\policy^{2} \in P^{2}} C(\policy^{1}, \policy^{2,*}) \leq C(\policy^{1,*}, \policy^{2,*}) \leq \inf_{\policy^{1} \in P^{1}} C(\policy^{1}, \policy^{2,*}).\] If such an equilibrium policy pair \((\policy^{1,*}, \policy^{2,*})\) exists, \(v^{*} = C(\policy^{1,*}, \policy^{2,*})\) is the \textit{equilibrium value} of the game.

\subsubsection{Reachability Objective}
 The event of eventually reaching set \(D\) is denoted by \(\lozenge D\). Under policies \(\policy^{\playerone}\) and \(\policy^{\playertwo}\), the probability of reaching set \(D\) from state \(\genericstate\) is denoted by \(\Pr^{\policy^{\playerone}, \policy^{\playertwo}}(\lozenge D | \genericstate)\). The probability of reaching set \(D\) from state \(\genericstate\) in \(L\) time steps is denoted by \(\Pr^{\policy^{\playerone}, \policy^{\playertwo}}(\lozenge_{\leq L} D | \genericstate)\).

  \(\winningstates\) denotes the set of winning states for Player \(\playerone\) for the reachability objective. Player \(\playerone\) \textit{wins} if and only if the game run \(\gamma = s_{0}a^{\playerone}_{0}a^{\playertwo}_{0}s_{1}a^{\playerone}_{1} a^{\playertwo}_{1}  \ldots\) satisfies \(s_{t} \in \winningstates\) for some \(t \geq 0\), i.e., \(\gamma\) satisfies \(\lozenge \winningstates\) .  A policy \(\policy^{\playerone}\) for Player \(\playerone\) is \textit{winning} if  \(\min_{\policy^{\playertwo} \in \Policy^{\playertwo}} {\Pr}^{\policy^{\playerone}, \policy^{\playertwo}}(\lozenge \winningstates | \initialstate) =1\). We denote the set of winning policies for Player \(\playerone\) by \(\Policy^{\playerone, win} \), and we denote the set of winning stationary policies for Player \(\playerone\) by \(\Policy^{\playerone, St,  win} = \Policy^{\playerone, St} \cap \Policy^{\playerone, win} \). For simplicity of presentation, we assume that all winning states are absorbing.

\subsection{Kullback--Leibler (KL) Divergence}
The support of a discrete probability distribution \(Q\) is denoted by \(Supp(Q)\). For discrete probability distributions \(Q_1\) and \(Q_2\) where \(Supp(Q_{1}) = \mathcal{X}\), the \textit{Kullback--Leibler (KL) divergence} between \(Q_1\) and \(Q_2\) is \(KL(Q_1 || Q_2) = \sum_{x \in \mathcal{X}} Q_1(x) \log \left( {Q_1(x)}/{Q_2(x)}\right)\) where \(\log\) denotes the natural logarithm. We define \(0 \log(0/0) = 0.\)
Data processing inequality~\citep{cover2012elements} states that any transformation \(T:\mathcal{X} \to \mathcal{Y}\) satisfies \(KL(Q_{1} || Q_{2}) \geq KL(T(Q_{1}) || T(Q_{2})).\)

 Let \(\policy^{\playerone'}\) be a policy for Player \(\playerone\). Note that \[KL\big( \Gamma^{\policy^{\playerone}, \policy^{\playertwo}}_{t} || \Gamma^{\policy^{\playerone'}, \policy^{\playertwo}}_{t} \big) \leq KL\big( \Gamma^{\policy^{\playerone}, \policy^{\playertwo}}_{t+1} || \Gamma^{\policy^{\playerone'}, \policy^{\playertwo}}_{t+1}\big)\] due to the data processing inequality. We define \[KL\big( \Gamma^{\policy^{\playerone}, \policy^{\playertwo}} || \Gamma^{\policy^{\playerone'}, \policy^{\playertwo}} \big) = \lim_{t \to \infty} KL\big( \Gamma^{\policy^{\playerone}, \policy^{\playertwo}}_{t} || \Gamma^{\policy^{\playerone'}, \policy^{\playertwo}}_{t} \big).\]
The limit either converges or diverges to \(\infty\) due to monotonicity of \( KL\big( \Gamma^{\policy^{\playerone}, \policy^{\playertwo}}_{t} || \Gamma^{\policy^{\playerone'}, \policy^{\playertwo}}_{t} \big).\) We denote \( KL\big( \Gamma^{\policy^{\playerone}, \policy^{\playertwo}} || \Gamma^{\policy^{\playerone'}, \policy^{\playertwo}} \big)\) with \(KL( \policy^{\playerone}, \policy^{\playertwo} || \policy^{\playerone'}, \policy^{\playertwo} )\) for notational simplicity.
\section{Problem Statement} \label{section:problemstatement}
We consider a two-player stochastic game \(\mathcal{G} = (\states, \actions^{\playerone}, \actions^{\playertwo}, \probs, \initialstate)\). In this game, Player \(1\) aims to satisfy the reachability objective \(\lozenge \winningstates\), i.e., win the game. Player \(\playerone\) may be \textit{hostile}, i.e., it may aim to reach \(\winningstates\) for malicious purposes. For example, in the cyber interaction scenario shown in Figure \ref{fig:dosattack}, Player \(\playerone\) may be a client performing a denial-of-service attack. Player \(\playertwo\) is not aware of the identity of Player \(\playerone\). In the cyber interaction scenario, Player \(\playertwo\) represents the server and does not know whether the client is an attacker. When Player \(1\) is hostile, its goal is not to expose its identity while winning the game. Player \(\playertwo\), on the other hand, aims to detect the identity of Player \(\playerone\), i.e., determine whether Player \(\playerone\) is hostile, in addition to making Player \(\playerone\) lose the game. In the cyber interaction scenario, the goal of Player \(\playerone\), i.e., a hostile client, is to perform an attack while not being detected by the server, and the goal of Player \(\playertwo\), i.e., the server, is to provide service to well-meaning clients while identifying hostile clients. In this setting, we assume that both players have full information on the current state and full information on each other's previous actions. 

We consider an average player as the reference point to measure identity concealment. The average player's policy encodes the expected behavior of a non-hostile player interacting with Player \(\playertwo\), and can be used to measure how much Player \(1\)'s policy \(\policy^{\playerone}\) exposes its identity and how well Player \(\playertwo\)'s policy \(\policy^{\playertwo}\) distinguishes hostile agents. The average player's policy \(\policy^{\average}\) is not necessarily designed to win the game against Player \(\playertwo\), but the average player can accidentally win the game due to the stochasticity of the environment or its policy. For example, in the cyber interaction scenario, \(\policy^{\average}\) may represent the average behavior of non-hostile clients, i.e., real users, interacting with the server. These clients may cause a denial-of-service, but their goal is not necessarily to cause a breakdown. We assume that the average player's policy \(\policy^{\average}\) is common knowledge. We also have the following assumption which ensures the computational tractability of the problems to be proposed.
\begin{assum}\label{ass:stationarityofaverage}
	The average player's policy \(\policy^{\average}\) is stationary on the state space \(\states\), i.e., \(\policy^{\average} \in \Policy^{\playerone, St}\).
\end{assum}

 Because an average player can win the game with a positive probability, a win in the game does not immediately identify Player \(\playerone\) as hostile. Therefore, Player \(\playerone\) aims to make its win look accidental and indistinguishable from an average player's win. On the flip side, Player \(\playertwo\) aims to design its policy in a way that the identity concealment of Player \(1\) is minimized, i.e., an average player and a hostile Player \(1\) produce different game runs.

We define the identity exposure payoff at time \(t\) as the KL divergence between the action distribution \(\mu^{\playerone}_{t}(\history_{t})\) under Player \(1\)'s policy \(\policy^{\playerone}=\mu^{1}_0\mu^{1}_1\ldots\) and the action distribution \(\policy^{\average}(\genericstate_{t})\) under the average player's policy \(\policy^{\average}\). Formally, the payoff of Players \(1\) and \(2\) is
\begin{align} \label{eq:histdepcost}
    \payoff(\genericstate_{t}, \mu^{\playerone}_{t}(\history_{t}), \mu^{\playertwo}_{t}(\history_{t}))&:= KL(\mu^{\playerone}_{t}(\history_{t})|| \policy^{\average}(\history_{t})) 
\end{align} at time \(t\).
For clarity of presentation, we restrict the feasible policy spaces of Players \(1\) and \(2\) to stationary policies, i.e., \(\policy^{\playerone} \in \Policy^{\playerone, St}\) and \(\policy^{\playertwo} \in \Policy^{\playertwo, St}.\) In this case, the payoff of Players \(1\) and \(2\) is
\begin{align*}
    \payoff(\genericstate_{t}, \policy^{\playerone}(\genericstate_{t}), \policy^{\playertwo}(\genericstate_{t}))&= KL(\policy^{\playerone}(\genericstate_{t})|| \policy^{\average}(\genericstate_{t})) 
    \\
    &= \sum_{\genericaction^{\playerone} \in \actions^{\playerone}} \policy^{\playerone}(\genericstate_{t}, \genericaction^{\playerone}) \log\left(\frac{\policy^{\playerone}(\genericstate_{t}, \genericaction^{\playerone})}{\policy^{\average}(\genericstate_{t}, \genericaction^{\playerone})}\right)
\end{align*} at time \(t\). This payoff decreases when \(\policy^{\playerone}(\genericstate_{t})\) gets more similar to \(\policy^{\average}(\genericstate_{t})\). The payoff is \(0\) when the action distributions of Player \(1\) and the average player is the same, i.e., \(\policy^{\playerone}(\genericstate_{t})=\policy^{\average}(\genericstate_{t}).\)

\textcolor{mycolor2}{Using the hitting time \(\randomstoppingtime\) that is the first hitting time to the set \(\absorbingstates\) of absorbing states, the \textit{zero-sum identity concealment objective function} is \[C(\policy^{1}, \policy^{2}):=\mathbb{E}\left[\sum_{t=0}^{\randomstoppingtime} KL(\policy^{\playerone}(\genericstate_{t})|| \policy^{\average}(\genericstate_{t})) \right]\] where the expectation is over the randomness of policies, \(\policy^{1}\) and \(\policy^{2}\), and the dynamics \(\probs\) of the game. Player \(\playerone\) is the \textit{minimizer} and Player \(\playertwo\) is the \textit{maximizer} of the zero-sum objective: Hostile Player \(1\) aims to behave similar to the average players and Player \(2\) aims to distinguish hostile Player \(1\) from the average players. The zero-sum objective accounts for the total identity exposure of Player \(\playerone\) until the effective end of the game, i.e., reaching an absorbing state. We discuss the relationship of this objective function with hypothesis testing in Section \ref{section:hypotesting}. }

	We define an \textit{identity concealment game} \( \identityconcealmentgame =  (\states, \actions^{\playerone}, \actions^{\playertwo}, \allowbreak \probs, \initialstate, \winningstates, \policy^{\average})\) as a two-player stochastic game with objective functions \[C(\policy^{1}, \policy^{2}) \allowbreak= \mathbb{E}\left[\sum_{t=0}^{\randomstoppingtime} KL(\policy^{\playerone}(\genericstate_{t})|| \policy^{\average}(\genericstate_{t})) \right]\] and \[1- {\Pr}^{\policy^{\playerone}, \policy^{\playertwo}}(\lozenge \winningstates | \initialstate),\]
 where Player \(1\) is the minimizer and Player \(\playertwo\) is the maximizer for both functions. We remark that the game is not well-defined due to multiple objective functions, and one needs combine these objective functions to find optimal policies. \cite{chen2013stochastic} showed that when multiple objective functions are combined with a conjunction where each predicate is a threshold for an objective function, it is, in general, computationally hard to compute optimal policies. To bypass computational issues, we pose the following problem that constrains the value of the objective function \(1- \Pr^{\policy^{\playerone}, \policy^{\playertwo}}(\lozenge \winningstates | \initialstate)\) to \(0\), i.e., Player 1 must use a winning policy.
\begin{prob} \label{problem}
	For a given identity concealment game \(\identityconcealmentgame\), determine whether there exists an equilibrium pair \((\policy^{\playerone,*}, \policy^{\playertwo,*})\in \Policy^{\playerone, St, win} \times \Policy^{\playertwo, St}\) of policies such that 
	\[\sup_{\policy^{2} \in \Policy^{\playertwo, St}} C(\policy^{1}, \policy^{2,*})\leq C(\policy^{1,*}, \policy^{2,*})\]
	and 
		\[ C(\policy^{1,*}, \policy^{2,*}) \leq \inf_{\policy^{1} \in \Policy^{\playerone, St, win}} C(\policy^{1}, \policy^{2,*}).\]
\end{prob}

\begin{rem}
	We state that Player \(1\) aims to win with probability \(1\) and assume that there exists such a winning policy. If there is not such a policy, one can use the weighted zero-sum objective function \(C(\policy^{1}, \policy^{2}) + \alpha (1 - \Pr^{\policy^{\playerone}, \policy^{\playertwo}}( \lozenge \winningstates | \initialstate ))\) where \(\alpha \in \mathbb{R} \cup \lbrace \infty \rbrace\) is the weight of the winning objective.  When winning in the game is a hard constraint for Player \(\playerone\), i.e., when \(\alpha = \infty\), the weighted zero-sum objective function recovers Problem \ref{problem}.
\end{rem}

Player \(\playerone\) can achieve a lower value than the equilibrium value of the game if Player \(\playertwo\) uses a suboptimal non-equilibrium policy. In this case, the optimal policy of Player \(\playerone\) is not necessarily the equilibrium policy, and the hostile Player \(\playerone\) may need to learn Player \(\playertwo\)'s policy to synthesize the optimal policy. While it is possible to learn Player \(\playertwo\)'s policy with a high amount of exploration, learning in this way is undesirable in the described adversarial setting since naive exploration would quickly reveal the identity of the hostile Player \(\playerone\). Furthermore, Player \(\playerone\) may not be able to collect game runs by directly interacting with Player \(\playertwo\) and may only observe Player \(\playertwo\)'s interactions with average players. Hence, the hostile Player \(\playerone\)'s goal is to learn Player \(\playertwo\)'s policy from runs collected under the average player's policy and compute the optimal policy. We propose the following problem.
\begin{prob} \label{problem:learning}
		For an identity concealment game \(\identityconcealmentgame\), let \(\policy^{\playertwo,\circ}\) be Player \(\playertwo\)'s policy that is unknown a priori to Player \(1\) and \(\policy^{\playerone,\circ} = \arg\min_{\policy \in \Policy^{1, win}} C(\policy^{1}, \policy^{2,\circ}).\) Given \(\optimalitygap > 0\), \(\losingprobability \in [0,1]\), and \(\failureprob \in [0,1]\), find an algorithm that uses a finite number of runs that are collected only using the average player's policy so that the (potentially history-dependent) output policy \(\policy^{\playerone}\) of the algorithm satisfies \[C(\policy^{1}, \policy^{2,\circ}) \leq C(\policy^{1,\circ}, \policy^{2,\circ}) + \optimalitygap\] and \[{\Pr}^{\policy^{\playerone}, \policy^{\playertwo,\circ}}(\lozenge \winningstates | \initialstate) \geq 1- \losingprobability,\] with probability at least \(1- \failureprob\).
\end{prob}

\subsection{KL Divergence Payoffs and Hypothesis Testing}
\label{section:hypotesting}
The identity concealment game has a KL objective function $C(\policy^{1}, \policy^{2}) = \mathbb{E}\left[\sum_{t=0}^{\randomstoppingtime} \allowbreak KL(\policy^{\playerone}(\genericstate_{t})|| \policy^{\average}(\genericstate_{t})) \right]$ motivated by statistical hypothesis testing. 
The sum of KL stage payoffs is equal to the KL divergence between the probability distribution of runs under Player \(\playerone\)'s policy \(\policy^{\playerone}\) and Player \(\playertwo\)'s policy \(\policy^{\playertwo}\), and the probability distribution of runs under the average player's policy \(\policy^{\average}\) and Player \(\playertwo\)'s policy \(\policy^{\playertwo}\). Formally, as we explain later in the proof of Lemma \ref{proposition:prolonginginfinite}, we have \[\mathbb{E}\left[\sum_{t=0}^{\randomstoppingtime} KL(\policy^{\playerone}(\genericstate_{t})|| \policy^{\average}(\genericstate_{t})) \right] = KL\left(\policy^{\playerone}, \policy^{\playertwo} || \policy^{\average}, \policy^{\playertwo}\right).\] 

By Sanov's theorem,  \(\exp\left( -n KL\left(\policy^{\playerone}, \policy^{\playertwo} || \policy^{\average}, \policy^{\playertwo} \right) \right) \) measures the probability that \(n\) random game runs with a hostile Player \(\playerone\) occur under the average player's policy. Consequently, as the number \(n\) of game runs or \(KL\left(\policy^{\playerone}, \policy^{\playertwo} || \policy^{\average}, \policy^{\playertwo}\right)\) increases Player \(\playertwo\) is more likely to identify a hostile player. More formally, as \(n KL\left(\policy^{\playerone}, \policy^{\playertwo} || \policy^{\average}, \policy^{\playertwo} \right) \) increases, the accuracy of the likelihood-ratio test~\citep{hogg1977probability} increases. The goal of Player \(\playerone\) is thus to minimize \(KL\left(\policy^{\playerone}, \policy^{\playertwo} || \policy^{\average}, \policy^{\playertwo} \right)\), while the goal of Player \(\playertwo\) is to maximize this value.

\section{Equilibrium Policies for Identity Concealment Games}

\begin{figure}
	\centering
    \resizebox{0.9\columnwidth}{!}{
		\begin{tikzpicture}[node distance = 1.5cm]
		    \tikzset{
	->, 
	>=stealth', 
	initial text=$ $, 
}
		
		    \node[state, initial] (initialstate) {$s^{0}$};
		    \node[state] [above right=  1cm and 2cm of initialstate] (cliffstate) {$s^{1}$};
		    \node[state] [below right= 1cm and 2cm of initialstate] (exilestate) {$s^{2}$};
		    \node[state] [below=of exilestate] (potwinstate1) {$s^{3}$};
		    \node[state] [below=0.5 cm of potwinstate1] (potwinstate3) {$s^{8}$};
		    \node[state] [right=3cm of exilestate] (potwinstate2) {$s^{4}$};
		    \node[state] [above=of cliffstate] (trapstate1) {$s^{5}$};
		    \node[state] [right=3cm of trapstate1] (trapstate2) {$s^{6}$};
		    \node[state] [right=3cm of cliffstate] (winstate) {$s^{7}$};
			\draw 			
			(initialstate) edge node[fill=white, sloped] {\footnotesize $\lbrace b\rbrace, \lbrace x\rbrace, 1$} (cliffstate)
			(initialstate) edge node[fill=white, sloped] {\footnotesize $\lbrace b\rbrace, \lbrace  y\rbrace, 1$} (exilestate)
			(initialstate) edge[bend right=60] node[fill=white, sloped] {\footnotesize $\lbrace a \rbrace, \lbrace x\rbrace, 1$} (potwinstate1)
			(initialstate) edge[bend right=60] node[fill=white, sloped] {\footnotesize $\lbrace a \rbrace, \lbrace y\rbrace, 1$} (potwinstate3)
			(exilestate) edge[loop below] node[below] {\footnotesize $\lbrace a \rbrace, \lbrace x, y\rbrace, 1$} (exilestate)
			(exilestate) edge node[fill=white] {\footnotesize $\lbrace b \rbrace, \lbrace x, y\rbrace, 1$} (cliffstate)
			(cliffstate) edge node[fill=white] {\footnotesize $\lbrace a, b \rbrace, \lbrace x, y\rbrace, 0.5$} (trapstate1)
			(cliffstate) edge node[fill=white, sloped] {\footnotesize $\lbrace a, b \rbrace, \lbrace x, y\rbrace, 0.5$} (winstate)
			(trapstate1) edge[bend right=20] node[fill=white, sloped] {\footnotesize $\lbrace a, b \rbrace, \lbrace x\rbrace, 1$} (trapstate2)
			(trapstate2) edge[bend right=20] node[fill=white, sloped] {\footnotesize $\lbrace a, b \rbrace, \lbrace x, y\rbrace, 1$} (trapstate1)
			(trapstate1) edge[bend right=60] node[fill=white, sloped] {\footnotesize $\lbrace a, b \rbrace, \lbrace y\rbrace, 1$} (initialstate)
			(potwinstate1) edge[bend right=30] node[fill=white, sloped] {\footnotesize $\lbrace a,b \rbrace, \lbrace x, y\rbrace, 1$} (potwinstate2)
			(potwinstate3) edge[bend right=45] node[fill=white, sloped] {\footnotesize $\lbrace a \rbrace, \lbrace x, y\rbrace, 1$}
			(potwinstate2)
			(potwinstate3) edge[loop below] node[below] {\footnotesize $\lbrace b \rbrace, \lbrace x, y\rbrace, 1$} (potwinstate3)
			(potwinstate2) edge node[fill=white] {\footnotesize $\lbrace a \rbrace, \lbrace x, y\rbrace, 1$} (winstate)
			(potwinstate2) edge node[fill=white, sloped] {\footnotesize $\lbrace b \rbrace, \lbrace x, y\rbrace, 1$} (cliffstate)
			(winstate) edge[loop above] node[fill=white] {\footnotesize $\lbrace a, b \rbrace, \lbrace x, y\rbrace, 1$} (winstate);
		\end{tikzpicture}
		
    }
	\caption{A identity concealment game where the actions of Player $\playerone$ are $a$ and $b$, and the actions of Player $\playertwo$ are $x$ and $y$. Nodes are the states of the game. The initial state is $s^{0}$ and $s^{7}$ is the only winning state for Player $\playerone$. The average player's policy $\policy^{\average}$ takes actions $a$ and $b$ uniformly randomly at every state.
	A label $D^{\playerone},D^{\playertwo},p$ of a directed edge from $\genericstate$ to $\altstate$ means $\probs(\genericstate, \genericaction^{\playerone}, \genericaction^{\playertwo}, \altstate) = p$ for every $\genericaction^{\playerone} \in D^{\playerone}$ and $\genericaction^{\playertwo} \in D^{\playertwo}$. A stationary trapping policy for Player $\playertwo$ takes action $x$ at every state.}\label{figure:ICexample}
\end{figure}

\label{section:eqpolicies}
In this section, we prove the existence of an equilibrium for the identity concealment game and provide the optimality equations to compute it.

If there exists an equilibrium, and the equilibrium value for value for \(C(\policy^{1}, \policy^{2}) \) is infinite in Problem \ref{problem}, and all winning stationary policies are equally good for Player \(\playerone\). We mainly focus on the more interesting case that there exists a winning stationary policy \(\policy^{\playerone} \in \Policy^{\playerone, St, win}\)  such that \(\max_{\policy^{\playertwo} \in \Policy^{\playertwo, St}} \) \(C(\policy^{1}, \policy^{2}) < \infty\).

We define that action \(\genericaction^{\playerone}\) is \textit{permissible} for Player \(\playerone\) at state \(\genericstate\) if  \(\policy^{\average}(\genericstate,\genericaction^{\playerone}) > 0\). For example, all actions are permissible at every state for Player $\playerone$ for the identity concealment game given in Figure \ref{figure:ICexample}. Note that if Player \(\playerone\) takes an impermissible action with a positive probability, then  \(C(\policy^{1}, \policy^{2})\) is infinite, i.e., with a positive probability Player \(\playertwo\) is certain that Player \(\playerone\) is not an average player, since an event happens with a positive probability under Player \(\playerone\)'s policy and zero probability under the average player's policy. Note that removing the impermissible actions does not change the equilibrium value if the equilibrium value is finite.

Given that the equilibrium value is finite, without loss of generality, we assume that all impermissible actions are removed from all states.
\begin{assum} \label{assum:permissibleactions}
    Every available action is permissible for Player \(1\).
\end{assum}

To find equilibrium policies for Problem \ref{problem}, we first identify the states at which Player \(\playertwo\) can win the game with a positive probability. These states can be found with a procedure similar to solving reachability games on graphs~\citep{chatterjee2008algorithms}.

State \(\genericstate\) is a \textit{trap} state for Player \(\playerone\) if there exists a policy \(\policy^{\playertwo} \in \Policy^{\playertwo}\) that satisfies \({\Pr}^{\policy^{\average}, \policy^{\playertwo}}( \lozenge \winningstates | \genericstate) = 0\).  For example states $s^{5}$ and $s^{6}$ are trap states in Figure \ref{figure:ICexample}.  The set of trap states is denoted by \(\trapstates\). The set of trap states is easy to find: Since \(\policy^{\average}\) is stationary, it induces an MDP for Player \(\playertwo\) given the game. On this MDP, to find \(\policy^{\playertwo, trap}\), we solve a reach-avoid problem where the objective is to avoid the winning states \(\winningstates\) for Player \(\playertwo\). \textcolor{mycolor2}{There exists a stationary policy \(\policy^{\playertwo, trap} \in \Policy^{\playertwo, St}\) that minimizes \({\Pr}^{\policy^{\average}, \policy^{\playertwo}}(\lozenge \winningstates | \genericstate)\) for every \(\genericstate \in \states\)~ \citep{baier2008principles}.} Since the trapping policy minimizes the winning probability of an average player for every state, we have  \(\genericstate\in \trapstates\) if and only if \({\Pr}^{\policy^{\average}, \policy^{\playertwo, trap}}(\lozenge \winningstates | \genericstate) = 0 \).

\textcolor{mycolor2}{
Under Assumption \ref{assum:permissibleactions}, we have \({\Pr}^{\policy^{\playerone}, \policy^{\playertwo, trap}}(\lozenge \winningstates | \genericstate) = 0 \) for every \(\genericstate\in \trapstates\) and \(\policy^{\playerone}\in \Policy^{\playerone}\). To observe this, we consider two directed graphs. The policy pair \((\policy^{\average}, \policy^{\playertwo, trap})\) induces a Markov chain. Let \(G^{(\policy^{\average}, \policy^{\playertwo, trap})}\) be a directed graph that represents the feasible transitions on this Markov chain. In this directed graph the states \(\trapstates\) are not connected to \(\winningstates\) since \({\Pr}^{\policy^{\average}, \policy^{\playertwo, trap}}(\lozenge \winningstates | \genericstate) = 0 \) for every \(\genericstate\in \trapstates\). Similarly, the policy pair \((\policy^{\playerone}, \policy^{\playertwo, trap})\) induces another Markov chain. Let \(G^{(\policy^{\playerone}, \policy^{\playertwo, trap})}\) be a directed graph that represents the feasible transitions on this Markov chain. Under Assumption \ref{assum:permissibleactions}, \((\policy^{\playerone}, \policy^{\playertwo, trap})\) must be a subgraph of \(G^{(\policy^{\average}, \policy^{\playertwo, trap})}\).
This is because \(\policy^{\average}\) takes every available action for Player \(1\) with a positive probability. Since \(G^{(\policy^{\playerone}, \policy^{\playertwo, trap})}\) is a subgraph of \(G^{(\policy^{\average}, \policy^{\playertwo, trap})}\), the states \(\trapstates\) are not connected to \(\winningstates\) in \(G^{(\policy^{\playerone}, \policy^{\playertwo, trap})}\). Hence, \({\Pr}^{\policy^{\playerone}, \policy^{\playertwo, trap}}(\lozenge \winningstates | \genericstate) = 0 \) for every \(\genericstate\in \trapstates\) and \(\policy^{\playerone}\in \Policy^{\playerone}\). 
}

\textcolor{mycolor2}{
A stationary winning policy \(\policy^{\playerone} \in\Policy^{\playerone, St, win}\) never visits a trap state regardless of Player \(\playertwo\)'s policy under Assumption \ref{assum:permissibleactions}. We show this by a contradiction. Consider policies \(\policy^{\playerone} \in\Policy^{\playerone, St, win}\) and \(\policy^{\playertwo} \in \policy^{\playertwo, St}\) that reach a trap state with a positive probability from the initial state, i.e., \(\Pr^{\policy^{\playerone}, \policy^{\playertwo}}(\lozenge \states^{trap} | \initialstate) > 0\). Consider a policy \(\policy^{\playertwo'}\) that is the same as \(\policy^{\playertwo}\) for all states in \(\states\setminus\states^{trap}\) and is the same as \(\policy^{\playertwo, trap}\) for all states in \(\states^{trap}\). We have \(\Pr^{\policy^{\playerone}, \policy^{\playertwo'}}(\lozenge \states^{trap} | \initialstate) = \Pr^{\policy^{\playerone}, \policy^{\playertwo}}(\lozenge \states^{trap} | \initialstate) > 0\) since \(\policy^{\playertwo'}\) is the same as \(\policy^{\playertwo}\) for all states in \(\states\setminus\states^{trap}\). We also have \(\Pr^{\policy^{\playerone}, \policy^{\playertwo'}}(\lozenge \winningstates | \genericstate) = 0\) for all \(\genericstate \in \states^{trap}\) since a policy \(\policy^{\playertwo'}\) is the same as \(\policy^{\playertwo, trap}\) for all states in \(\states^{trap}\). Hence, we have \(\Pr^{\policy^{\playerone}, \policy^{\playertwo'}}(\lozenge \winningstates | \initialstate) < 1\) which contradicts with the fact that \(\policy^{\playerone}\) is a winning policy. Without Assumption \ref{assum:permissibleactions}, there could be a \(\policy^{\playerone} \in\Policy^{\playerone, St, win}\) that visits a trap state. All such policies take an impermissible action with a positive probability and yield an infinite objective value.
}

We find the set \(\potentiallywinningstates\) of potentially winning states for which there exists a policy \(\policy^{\playerone}\) for Player \(\playerone\) that reaches \(\winningstates\) with probability \(1\) for all \(\policy^{\playertwo} \in \Policy^{\playertwo}\) and avoids \(\trapstates\). For example, $s^{0}$, $s^{3}$, $s^{4}$, $s^{7}$, and $s^{8}$ are the potentially winning states in Figure \ref{figure:ICexample}. We remark that there might be some states from which Player $\ref{figure:ICexample}$ can avoid the trap states with probability $1$, but never reach the winning states. For example, $s^{2}$ is such a state in Figure \ref{figure:ICexample}. The set \(\potentiallywinningstates\) of potentially winning states can be found by iteratively expanding  \(\winningstates\) as in the attractor computation for two-player reachability games~\citep{chatterjee2008algorithms}. \textcolor{mycolor2}{We note that stationary policies for Player \(1\) suffice to achieve maximal \(\potentiallywinningstates\) against all possible policies of Player \(\playertwo\) since the game has the Markov property.} If a pair of equilibrium policies exist, then only the states in \(\potentiallywinningstates\) are visited with a positive probability since from all states in \(\states \setminus \potentiallywinningstates\), there exists a policy for Player \(\playertwo\) such that \(\winningstates\) is reached with a probability strictly less than \(1\).  We define that at state \(\genericstate \in \potentiallywinningstates\) action \(\genericaction^{\playerone}\) is \textit{safe} for Player \(\playerone\) if and only if all potential successor states are in \(\potentiallywinningstates\), i.e.,  \(\probs(\genericstate,\genericaction^{\playerone},\genericaction^{\playertwo}, \altstate) = 0\) for all \(\genericaction^{\playertwo} \in \actions^{\playertwo}\) and \(\altstate \in \states \setminus \potentiallywinningstates\). Note that for every state in \(\potentiallywinningstates\), there exists a safe action due to the construction of \(\potentiallywinningstates\). For example, $a$ and $b$ are safe actions for states $s^{3}$,$s^{7}$, and $s^{8}$, and $a$ is the only safe action for states $s^{0}$ and $s^{4}$.

Having identified the set of states from which Player $\playerone$ can win the game with probability $1$, we now focus on the existence of equilibrium policies. We note that the stage payoff $\sum_{\genericaction^{\playerone} \in \actions^{1}} \policy^{\playerone}(\genericstate_{t}, \genericaction^{\playerone}) \allowbreak  \log \left( \frac{\policy^{\playerone}(\genericstate_{t}, \genericaction^{\playerone}) }{\policy^{\average}(\genericstate_{t}, \genericaction^{\playerone})} \right)$ is a convex function of the policy parameters of the minimizer, i.e., Player $\playerone$ and a concave function of the policy parameters of the maximizer, i.e., Player $\playertwo$. Zero-sum stochastic games with such payoffs have an equilibrium when the payoffs are discounted~\citep{bacsar1998dynamic}. However, the game that we consider does not have a discount. To show the existence of an equilibrium, we need to prove additional properties of the identity concealment game.

\textcolor{mycolor2}{Lemma \ref{proposition:prolonginginfinite} shows that if the initial state is in the potentially winning states, then there exists a (stationary) policy that makes the KL objective function finite. Furthermore,  the occupancy measure at all states, but the states in \(\absorbingstates\) must be finite in order to have a finite value for the KL objective function. To show this, we consider the visitation distributions for an arbitrary state since the KL divergence between the visitation distributions is a lower bound on the KL objective function. Proof of Lemma \ref{proposition:prolonginginfinite} shows that the policy pairs that induce infinite occupancy measure for a state that is not in \(\absorbingstates\) lead to a visitation distribution that has infinite KL divergence from the distribution under the average player's policy.} Formally, a pair \((\policy^{\playerone}, \policy^{\playertwo})\) of (history-dependent) policies is \textit{prolonging} if \(\sum_{t=0}^{\infty} \Pr^{\policy^{\playerone}, \policy^{\playertwo}}(\genericstate_{t} = \genericstate|\genericstate_{0}) = \infty \) for some \(\genericstate \in \potentiallywinningstates \setminus \absorbingstates\). All prolonging pairs \((\policy^{\playerone}, \policy^{\playertwo})\) of policies satisfy \(C(\policy^{1}, \policy^{2}) = \infty\). For example, consider state $s^{8}$ of the identity concealment game shown in Figure \ref{figure:ICexample}. There exists history-dependent policies for Player \(1\) that induces \(\sum_{t=0}^{\infty} \Pr^{\policy^{\playerone}, \policy^{\playertwo}}(\genericstate_{t} = \genericstate|\genericstate^{0}) = \infty \) if \(s^{8}\) is reached with a positive probability. All such policy pairs have \(C(\policy^{1}, \policy^{2}) = \infty\). We use properties given in Lemma \ref{proposition:prolonginginfinite} to show the existence of an equilibrium.

\begin{lem} \label{proposition:prolonginginfinite}
	If \(\initialstate \in \potentiallywinningstates\), then there exists a winning policy \(\policy^{\playerone, fin} \in \Policy^{\playerone, win}\) that satisfies \(C(\policy^{1,fin}, \policy^{2}) < \infty\), and \(\sum_{t=0}^{\infty} \Pr^{\policy^{\playerone, fin}, \policy^{\playertwo}}(\genericstate_{t} = \genericstate | \genericstate_{0}) < \infty \) for all \(\genericstate \in \potentiallywinningstates \setminus \winningstates\) and \(\policy^{\playertwo} \in \Policy^{\playertwo}\).
 
 If \(\sum_{t=0}^{\infty} \Pr^{\policy^{\playerone, inf}, \policy^{\playertwo}}(\genericstate_{t} = \genericstate|\genericstate_{0}) = \infty\) for some \(\genericstate \in \potentiallywinningstates \setminus \winningstates\) and \(\policy^{\playerone, inf} \in \Policy^{\playerone}\), then \(C(\policy^{1,inf}, \policy^{2}) = \infty\).
\end{lem}

 We use the additional Lemma \ref{lemma:ultimatelemma} to prove Lemma \ref{proposition:prolonginginfinite}. The proof of Lemma \ref{lemma:ultimatelemma} follows from that \(\sum_{n \in C'} \mathcal{D}^{1}(n) n = \infty\) where \(n \in C'\) if and only if \(\mathcal{D}^{1}(n) > c_{1} \exp(-n c_{2}/2)\).
\begin{lem} \label{lemma:ultimatelemma}
	Let \(\mathcal{D}^{1}\) and \(\mathcal{D}^{2}\) be discrete probability distributions such that \(Supp(\mathcal{D}^{1}),Supp(\mathcal{D}^{2}) \subseteq \mathbb{N}\). If \(\sum_{n = 0}^{\infty} \mathcal{D}^{1}(n) n = \infty\) and there exist  \(c_{1}, c_{2} \in (0 , \infty)\) such that \(\mathcal{D}^{2}(n) \leq c_{1}\exp(-c_{2}n)\), then \(KL(\mathcal{D}^{1} || \mathcal{D}^{2}) = \infty\).
\end{lem}

\begin{proof}[Proof of Lemma \ref{proposition:prolonginginfinite}]
     We prove the first part of the lemma by constructing a policy using safe actions and the definition of potentially winning states. For the second part, we lower bound the KL objective using the data processing inequality, and the time distributions at the states in \(\potentiallywinningstates \setminus \states^{R}\). We show that the lower bound and, consequently, the objective function are infinite if the Player \(\playerone\)'s policy has infinite occupancy measure at \(\potentiallywinningstates \setminus \winningstates\).

    We show the existence of a stationary \(\policy^{\playerone, fin}\) by construction. At states in \( \potentiallywinningstates\), \(\policy^{\playerone, fin}\) takes all permissible, safe actions uniformly randomly. To show \(C(\policy^{1,fin}, \policy^{2}) < \infty\), we first note that by definition of conditional expectation, \(C(\policy^{1,fin}, \policy^{2}) =  \sum_{\genericstate \setminus \absorbingstates}   \sum_{t = 0}^{\infty}  \allowbreak {\Pr}^{\policy^{\playerone, fin}, \policy^{\playertwo}}(\genericstate_{t} = \genericstate|\initialstate) KL(\policy^{\playerone}(\genericstate)|| \policy^{\average}(\genericstate))\).
    
    For every \(\genericstate \in \potentiallywinningstates\), we have \(KL(\policy^{\playerone}(\genericstate)|| \policy^{\average}(\genericstate)) \allowbreak < \allowbreak \infty\) since \(\policy^{\playerone, fin}\) takes only permissible actions. Let \(\bar{c} = \max_{s \in \potentiallywinningstates} KL(\policy^{\playerone}(\genericstate)|| \policy^{\average}(\genericstate))\). 
    
    By definition of \(\potentiallywinningstates\), there must exist a state \(\genericstate_{t} \in \potentiallywinningstates\) and such that \({\Pr}^{\policy^{\playerone, fin}, \policy^{\playertwo}}(\genericstate_{t+1} \in \winningstates | \genericstate_{t}) > 0\) for all \(\policy^{\playertwo}\). Similarly, \({\Pr}^{\policy^{\playerone, fin}, \policy^{\playertwo}}(\lozenge_{\leq \cardstates} \winningstates| \history_{t}) > 0\) for all \(\history_{t} \in \histories_{t}\). Since the game ends in every \(\cardstates\) steps with a positive probability, we have \(\sum_{\genericstate \in \states\setminus \absorbingstates}   \sum_{t = 0}^{\infty} {\Pr}^{\policy^{\playerone, fin}, \policy^{\playertwo}}(\genericstate_{t} = \genericstate |\initialstate) \leq \bar{t}< \infty\). Therefore, we have \(C(\policy^{1,fin}, \policy^{2}) \leq \bar{c} \bar{t} \allowbreak < \allowbreak \infty.\)

	We now prove that if \(\sum_{t=0}^{\infty} {\Pr}^{\policy^{\playerone, inf}, \policy^{\playertwo}}(\genericstate_{t} = \genericstate) = \infty\) for some \(\genericstate \in \potentiallywinningstates \setminus \winningstates\), then \(C(\policy^{1,inf}, \policy^{2})\) is infinite. 
We first represent the objective function \(C(\policy^{1}, \policy^{2})\) as the KL divergence between the probability distribution of runs under Player \(\playerone\)'s policy \(\policy^{\playerone}\) and Player \(\playertwo\)'s policy \(\policy^{\playertwo}\), and the probability distribution of runs under the average player's policy \(\policy^{\average}\) and Player \(\playertwo\)'s policy \(\policy^{\playertwo}\). In detail, 
	\begin{linenomath*}
		\begin{subequations}
		\begin{align*} 
		&C(\policy^{1}, \policy^{2}) =\mathbb{E} \left[ \sum_{t = 0}^{\randomstoppingtime} \sum_{\genericaction^{\playerone} \in \actions^{1}} \mu^{\playerone}_{t}(\history_{t}, \genericaction^{\playerone})  \log \left( \frac{\mu^{\playerone}_{t}(\history_{t}, \genericaction^{\playerone}) }{\policy^{\average}(\genericstate_{t}, \genericaction^{\playerone})} \right)   \right]
				\\
    &=\mathbb{E} \left[ \sum_{t = 0}^{\infty} \sum_{\genericaction^{\playerone} \in \actions^{1}} \mu^{\playerone}_{t}(\history_{t}, \genericaction^{\playerone})  \log \left( \frac{\mu^{\playerone}_{t}(\history_{t}, \genericaction^{\playerone}) }{\policy^{\average}(\genericstate_{t}, \genericaction^{\playerone})} \right)   \right]
				\\
		&= \mathbb{E} \left[ \sum_{t = 0}^{\infty} \sum_{\substack{\genericstate_{t+1} \in  \states \\ \genericaction^{\playerone} \in \actions^{1}\\\genericaction^{\playertwo} \in \actions^{\playertwo}}} \mu^{\playerone}_{t}(\history_{t}, \genericaction^{\playerone}) \mu^{\playertwo}_{t}(\history_{t}, \genericaction^{\playertwo}) \probs(\genericstate_{t},\genericaction^{\playerone},\genericaction^{\playertwo}, \genericstate_{t+1}) \right.
		\\
		& \left. \log \left( \frac{\mu^{\playerone}_{t}(\history_{t}, \genericaction^{\playerone}) \mu^{\playertwo}_{t}(\history_{t}, \genericaction^{\playertwo}) \probs(\genericstate_{t},\genericaction^{\playerone},\genericaction^{\playertwo}, \genericstate_{t+1}) }{\policy^{\average}(\genericstate_{t}, \genericaction^{\playerone}) \mu^{\playertwo}_{t}(\history_{t}, \genericaction^{\playertwo}) \probs(\genericstate_{t},\genericaction^{\playerone},\genericaction^{\playertwo}, \genericstate_{t+1})} \right)  \right] 
		\\
		&=\lim_{t \to \infty} \sum_{\gamma_{t} \in Supp(\Gamma^{\policy^{\playerone}, \policy^{\playertwo}}_{t})}  {\Pr}^{\policy^{\playerone}, \policy^{\playertwo}} (\gamma_{t}) \log\left( \frac{{\Pr}^{\policy^{\playerone}, \policy^{\playertwo}} (\gamma_{t})}{{\Pr}^{\policy^{\average}, \policy^{\playertwo}} (\gamma_{t})} \right)
		\end{align*}
	\end{subequations} 
	\end{linenomath*}
	where the first equality is by definition, \textcolor{mycolor2}{the second equality is because for \(t \geq \randomstoppingtime\) \(\genericstate_{t} \in \absorbingstates\) and every state \(\genericstate \in \absorbingstates\) has a single action inducing \(0\) KL divergence cost,} and the last inequality is due to the chain rule of KL divergence and Markovianity of the game.

Let \(\lozenge \bigcirc D\) denote the event of eventually reaching set \(D\) starting from the next time step. The game run \(\gamma = s_{0}a^{\playerone}_{0}a^{\playertwo}_{0}s_{1}a^{\playerone}_{1} a^{\playertwo}_{1}  \ldots\) satisfies \(\lozenge \bigcirc D\) if and only if there exists \(s_{t} \in D\) for some \(t \geq 1\).
 We first claim that for all \(\policy^{\playertwo} \in \Policy^{\playertwo}\), \(t \geq 0\), and \(\history_{t} \in \histories_{t}\) such that \( \genericstate_{t} \in \potentiallywinningstates \setminus \winningstates\), we have  \({\Pr}^{\policy^{\average}, \policy^{\playertwo}}(\lozenge \bigcirc \lbrace \genericstate_{t} \rbrace | \history_{t}) \allowbreak < 1 \). In words, a state in \(\potentiallywinningstates \setminus \winningstates\) will not be visited back with a positive probability. Note that stationary policies for Player \(\playertwo\) suffice to maximize the reachability probability to a state in the MDP induced by \(\policy^{\average}\). If \({\Pr}^{\policy^{\average}, \policy^{\playertwo}}(\lozenge \bigcirc \lbrace \genericstate_{t} \rbrace | \history_{t}) \allowbreak = 1 \) for some \(\policy^{\playertwo} \in \Policy^{\playertwo, St}\), then we have \({\Pr}^{\policy^{\average}, \policy^{\playertwo}}(\lozenge \winningstates | \genericstate_{t}) = 0\) and \(\genericstate_{t}\) must be a trap state. This yields a contradiction since \(\genericstate \in \potentiallywinningstates\) and \(\potentiallywinningstates \cap \trapstates = \emptyset\). Thus, for all \(\policy^{\playertwo} \in \Policy^{\playertwo}\), \(t \geq 0\), and \(\history_{t} \in \histories_{t}\) such that \(\genericstate_{t} \in \potentiallywinningstates \setminus \winningstates\), we have \({\Pr}^{\policy^{\average}, \policy^{\playertwo}}(\lozenge \bigcirc \lbrace \genericstate_{t} \rbrace | \history_{t}) \allowbreak < 1 \). 
	Let \(N^{\average, \playertwo}_{\genericstate}\) be a random variable denoting the number of times that \(\genericstate \in \potentiallywinningstates \setminus \winningstates\) appears in a random run under \(\policy^{\average}\) and \(\policy^{\playertwo}\). Similarly, let \(N^{\playerone, \playertwo}_{\genericstate}\) be a random variable denoting the number of times that \(\genericstate \in \potentiallywinningstates \setminus \winningstates\) appears in a random run under \(\policy^{\playerone}\) and \(\policy^{\playertwo}\). Since \({\Pr}^{\policy^{\average}, \policy^{\playertwo}}(\lozenge \bigcirc \lbrace \genericstate \rbrace | \history_{t}, s_{t} = s) \allowbreak <  1 \), there exists a \(c \in [0,1)\) such that \(\Pr(N^{\average, \playertwo}_{\genericstate} = k) \leq  c^{k-1}\) for all \(k \geq 1\) and \(\Pr(N^{\average, \playertwo}_{\genericstate} = 0) \leq  1\). If \(\sum_{t=0}^{\infty} \Pr^{\policy^{\playerone, inf}, \policy^{\playertwo}}(\genericstate_{t} = \genericstate) = \infty\) for some \(\genericstate \in \potentiallywinningstates \setminus \winningstates\), then by Lemma \ref{lemma:ultimatelemma}, we have  \(KL(N^{\playerone, \playertwo}_{\genericstate} || N^{\average, \playertwo}_{\genericstate}) = \infty\) for some \(\genericstate \in \potentiallywinningstates \setminus \winningstates\). By the data processing inequality, for all \(\genericstate \in S\) we have \(KL(N^{\playerone, \playertwo}_{\genericstate} || N^{\average, \playertwo}_{\genericstate}) \leq  KL(\policy^{\playerone,inf}, \policy^{\playertwo} ||\policy^{\average,inf}, \policy^{\playertwo}).\) Thus,  \(KL(\policy^{\playerone,inf}, \policy^{\playertwo} ||\policy^{\average}, \policy^{\playertwo}) = C(\policy^{1,inf}, \policy^{2}) = \infty\).
\end{proof}

\textcolor{mycolor2}{
We note three facts: \(1)\) All prolonging pairs \((\policy^{inf, \playerone}, \policy^{\playertwo})\) of policies for which \(\sum_{t=0}^{\randomstoppingtime} \Pr^{\policy^{inf, \playerone}, \policy^{\playertwo}}(\genericstate_{t} = \genericstate) = \infty \) for some \(\genericstate \in \potentiallywinningstates \setminus \winningstates\),  satisfy \(C(\policy^{1,inf}, \policy^{2}) = \infty\), \(2)\) There exists a winning policy \(\policy^{\playerone, fin}\) for Player \(\playerone\) that has \(C(\policy^{1, fin}, \policy^{2}) < \infty\), and \(3)\) The payoff for each time step is a convex, continuous function of Player \(\playerone\)'s action distribution and a concave, continuous function of Player 2's action distribution. Due to these facts it is sufficient to consider only the stationary policies to find an equilibrium policy pair~\citep{patek1999stochastic}. When these conditions hold, Bellman's optimality equation leads to a unique fixed point, and there exists a stationary policy for a player that induces the optimal set of occupancy measures and achieves Bellman optimality when the other player's policy is fixed\footnote{\textcolor{mycolor2}{The game model in \citep{patek1999stochastic} has metric action spaces and the convex-concave property of the payoff function is not needed. We consider a game model with a finite set of actions and need the convex-concave property to show that the deterministic policies are sufficient for the equivalent game model with the metric action spaces. Consequent the randomized policies are sufficient for the game model that we consider. We discuss this implication in the proof of Proposition \ref{prop:eqpolicies}.}}.}
\begin{prop} \label{prop:eqpolicies}
	For an identity concealment game \(\identityconcealmentgame\), if there exists a winning policy \(\policy^{\playerone}\) for Player \(\playerone\) for which \(\Pr^{\policy^{\playerone}, \policy^{\playertwo}}(\lozenge \winningstates | \initialstate) = 1\) for all \(\policy^{\playertwo} \in \Policy^{\playertwo}\), then there exists an equilibrium pair \((\policy^{\playerone,*}, \policy^{\playertwo,*})\in \Policy^{\playerone, St, win} \times \Policy^{\playertwo, St}\) of policies such that 	\[\sup_{\policy^{2} \in \Policy^{\playertwo, St}} C(\policy^{1}, \policy^{2,*})\leq C(\policy^{1,*}, \policy^{2,*})\]
	and 
		\[ C(\policy^{1,*}, \policy^{2,*}) \leq \inf_{\policy^{1} \in \Policy^{\playerone, St, win}} C(\policy^{1}, \policy^{2,*}).\]
\end{prop}
\color{black}
\begin{proof}[Proof of Proposition \ref{prop:eqpolicies}]
    We consider two cases, the equilibrium value for the KL objective is finite and infinite. 
    
	We first consider the finite case. The existence of an equilibrium follows from the conditions given in \citep{patek1999stochastic}: \(1)\) All prolonging pairs \((\policy^{inf, \playerone}, \policy^{\playertwo})\) of policies for which \(\sum_{t=0}^{\randomstoppingtime} \Pr^{\policy^{inf, \playerone}, \policy^{\playertwo}}(\genericstate_{t} = \genericstate) = \infty \) for some \(\genericstate \in \potentiallywinningstates \setminus \winningstates\),  satisfy \(C(\policy^{1,inf}, \policy^{2}) = \infty\), \(2)\) There exists a winning policy \(\policy^{\playerone, fin}\) for Player \(\playerone\) that has \(C(\policy^{1, fin}, \policy^{2}) < \infty\), and \(3)\) The payoff for each time step is a convex function of Player \(\playerone\)'s action distribution and a concave function of Player 2's action distribution. We show that these conditions hold for the identity concealment games and prove the existence of an equilibrium.
 
 Without loss of generality, assume that Player \(\playerone\) only takes actions that are safe. Let Player \(\playerone\)'s actions \(\actions^{\playerone}(\genericstate)\) be enumerated from \(1\) to \(|\actions^{\playerone}|\) and Player 2's actions \(\actions^{\playertwo}\) be enumerated from \(1\) to \(|\actions^{\playertwo}|\) for all \(\genericstate \in\states\). Denote the \(n\)-dimensional probability simplex by \(\simplex^{n}\).
	
	Define a zero-sum two-player stochastic game \(\hat{\mathcal{G}} = (\states, \hat{\actions}^{1}, \hat{\actions}^{\playertwo}, \hat{\probs}, \initialstate, \winningstates)\) with compact action spaces  where \(\hat{\actions}^{\playerone}\) and \(\hat{\actions}^{\playertwo}\) are metric set of actions for Players \(\playerone\) and 2, respectively. Player \(\playerone\) and 2's policies are \(\hat{\policy}^{\playerone}\) and \(\hat{\policy}^{\playertwo}\), respectively. At time \(t\), Player \(\playerone\)'s decision function is \(\hat{\mu}^{\playerone}_{t}: \histories_{t} \to \simplex^{|\actions^{\playerone}|}\) and Player 2's decision function is  \(\hat{\mu}^{\playertwo}_{t}: \histories_{t} \to \simplex^{|\actions^{\playertwo}|}\). Player \(\playerone\) and 2's feasible policies are \(\hat{\Policy}^{\playerone}\) and \(\hat{\Policy}^{\playertwo}\), respectively. Let \(\hat{\Policy}^{1, win}\) be the set of winning policies for Player 1 such that \(\hat{\Policy}^{1, win} = \left\lbrace \policy^{\playerone} |  \min_{\hat{\policy}^{\playertwo} \in \hat{\Policy}^{\playertwo}} {\Pr}^{\hat{\policy}^{\playerone}, \hat{\policy}^{\playertwo}}(\lozenge \winningstates | \initialstate) =1 \right\rbrace. \) We define \(\hat{\actions}^{\playerone}, \hat{\actions}^{\playertwo},\) and \(\hat{\probs}\) such that the following is satisfied: \[\hat{\probs}(\genericstate, \hat{\genericaction}^{\playerone}, \hat{\genericaction}^{\playertwo}, \altstate) = \sum_{i=1}^{|\actions^{\playerone}|} \sum_{j=1}^{|\actions^{\playertwo}|} \hat{\genericaction}^{\playerone}(i) \hat{\genericaction}^{\playertwo}(j) \probs(\genericstate, i, j, \altstate)\] for all \(\genericstate \in \states\),  \(\hat{\genericaction}^{\playerone} \in \hat{\actions}^{\playerone} = \simplex^{|\actions^{\playerone}|}\), \(\hat{\genericaction}^{\playertwo} \in \hat{\actions}^{\playertwo} = \simplex^{|\actions^{\playertwo}|}\), and \(\altstate \in \states\). Define payoff function \(\hat{c}(\genericstate, \hat{\genericaction}^{\playerone}, \hat{\genericaction}^{\playertwo}) = \sum_{\genericaction^{\playerone} \in \actions^{\playerone}} \) \(\hat{\genericaction}^{\playerone}(i)\) \( \log\left({\hat{\genericaction}^{\playerone}(i)}/{ \policy^{\average}(\genericstate,\genericaction^{\playerone})} \right)\) for all \(\genericstate \in \potentiallywinningstates\setminus \winningstates\), \(\hat{c}(\genericstate, \hat{\genericaction}^{\playerone}, \hat{\genericaction}^{\playertwo}) = 0\) for all \(\genericstate \in \winningstates\), and \(\hat{c}(\genericstate,\hat{\genericaction}^{\playerone}, \hat{\genericaction}^{\playertwo}) = \infty\) for all \(\genericstate \in \states \setminus \potentiallywinningstates\). We consider \(\hat{\mathcal{G}}\) with the objective function \(\hat{C}(\hat{\policy}^{1}, \hat{\policy}^{2}) = \mathbb{E}^{\hat{\policy}^{\playerone}, \hat{\policy}^{\playertwo}}\left[\sum_{t=0}^{\randomstoppingtime} \hat{c}(\genericstate, \hat{\genericaction}^{\playerone}_{t}, \hat{\genericaction}^{\playertwo}_{t})\right]\) where Player \(\playerone\) is the minimizer and Player 2 is the maximizer. Note that the payoff function is a convex function of \(\hat{\genericaction}^{\playerone}\) and a concave function of \(\hat{\genericaction}^{\playertwo}\). We also note that by definition \(\hat{C}(\hat{\policy}^{1}, \hat{\policy}^{2})=\mathbb{E}^{\hat{\policy}^{\playerone}, \hat{\policy}^{\playertwo}}\left[\sum_{t=0}^{\randomstoppingtime} \hat{c}(\genericstate, \hat{\genericaction}^{\playerone}_{t}, \hat{\genericaction}^{\playertwo}_{t})\right]\) is equal to the value of \(C(\policy^{1}, \policy^{2}) = \mathbb{E}\left[\sum_{t=0}^{\randomstoppingtime} KL(\mu^{\playerone}_{t}(\genericstate_{t})|| \policy^{\average}(\genericstate_{t})) \right]\) if for all \(\genericstate \in \states\), \(t \geq 0\), we have \( \hat{\mu}^{\playerone}_{t}= [\mu^{\playerone}_{t}(\genericstate, 1), \ldots, \mu^{\playerone}_{t}(\genericstate, |\actions^{\playerone}|)]\) and \( \hat{\mu}^{\playertwo}_{t}= [\mu^{\playertwo}_{t}(\genericstate, 1), \ldots, \mu^{\playertwo}_{t}(\genericstate, |\actions^{\playertwo}|)]\). 
	
	Due to Lemma \ref{proposition:prolonginginfinite} and the above equivalence between the objective functions of \(\identityconcealmentgame\) and \(\hat{\mathcal{G}}\), all prolonging policy pairs \((\hat{\policy}^{\playerone, inf}, \hat{\policy}^{\playertwo})\) has an infinite objective value for Player \(\playerone\). Similarly, due to Lemma \ref{proposition:prolonginginfinite}, there exists a policy \(\hat{\policy}^{\playerone, fin}\) that incurs a finite objective value for Player \(\playerone\) for all policies of Player \(\playertwo\). Note that by construction \(\hat{\mathcal{G}}\) and \(\hat{\payoff}\), every policy pair \((\hat{\policy}^{\playerone}, \hat{\policy}^{\playertwo})\) with \(\hat{C}(\hat{\policy}^{1}, \hat{\policy}^{2}) < \infty\) reaches \(\winningstates\) with probability 1. Also note that there exists a \(\hat{\policy}^{2}\) for every \(\hat{\policy}^{1} \in \hat{\Policy}^{1} \setminus \hat{\Policy}^{1, win} \) that makes \(\hat{C}(\hat{\policy}^{1}, \hat{\policy}^{2}) = \infty\). Hence, we limit the feasible policies of Player \(1\) to \(\hat{\Policy}^{1, win}\).
	
	Since all prolonging policy pairs incur an infinite objective value for Player \(\playerone\) and there exists a policy \(\hat{\policy}^{\playerone, fin}\) that incurs a finite objective value for Player \(\playerone\), by Proposition 4.6 of \citep{patek1999stochastic}, the equilibrium value is unique and there exists an equilibrium policy pair for \(\hat{\mathcal{G}}\). Furthermore, there exists stationary pair \((\hat{\policy}^{\playerone, *}, \hat{\policy}^{\playertwo, *}) \in \hat{\Policy}^{\playerone, St, win} \times \hat{\Policy}^{\playertwo, St} \) of policies which achieve an equilibrium, i.e.,  \[\sup_{\hat{\policy}^{2} \in \hat{\Policy}^{\playertwo}} \hat{C}(\hat{\policy}^{1}, \hat{\policy}^{2}) \leq \hat{C}(\hat{\policy}^{1,*}, \hat{\policy}^{2,*}) \leq \inf_{\hat{\policy}^{1} \in \hat{\Policy}^{\playerone, win}} \hat{C}(\hat{\policy}^{1}, \hat{\policy}^{2,*}) .\]  We also note that the convexity of \(\hat{c}(\genericstate_{t}, \hat{\genericaction}^{\playerone}_{t}, \hat{\genericaction}^{\playertwo}_{t})\) in \(\hat{\genericaction}^{\playerone}\) and the concavity  in \(\hat{\genericaction}^{\playertwo}\) implies that the deterministic policies suffice for Player \(\playerone\) and Player 2 in \(\hat{\mathcal{G}}\).  Since there is a one-to-one mapping between the deterministic policies of \(\hat{\mathcal{G}}\) and all policies of \(\identityconcealmentgame\), there also exists an equilibrium stationary policy pair for \(\identityconcealmentgame\), i.e., there exists \((\policy^{\playerone,*}, \policy^{\playertwo,*})\in \Policy^{\playerone, St, win} \times \Policy^{\playertwo, St}\) such that \[\sup_{\policy^{2} \in \Policy^{\playertwo}} C(\policy^{1}, \policy^{2,*}) \leq C(\policy^{1,*}, \policy^{2,*}) \leq \inf_{\policy^{1} \in \Policy^{\playerone, win}} C(\policy^{1}, \policy^{2,*}).\] Restricting the policy spaces to stationary policies yields the desired result. Note that \(\winningstates\) is eventually reached with probability \(1\) under this equilibrium policy pair since the occupancy measures at \(\states \setminus \winningstates\) are finite.

    Finally, consider the infinite case. Since the KL objective function is infinite, we must have \(\initialstate \not \in \potentiallywinningstates\) due to Lemma \ref{proposition:prolonginginfinite}. Since there exists a stationary winning policy \(\policy^{\playerone}\), but \(\initialstate \not \in \potentiallywinningstates\), it implies that all winning policies take an impermissible action with a positive probability. Let stationary policy \(\policy^{\playertwo}\) for Player \(\playertwo\) be equal to \(\policy^{\playertwo, trap}\) for the states in \(\states \setminus \potentiallywinningstates\) and take actions uniformly randomly for the states in \(\potentiallywinningstates\). Every \(\policy^{\playerone} \in \Policy^{1, win}\) has \(C(\policy^{1}, \policy^{2})  = \infty\) and \(\Pr^{\policy^{\playerone}, \policy^{\playertwo}}(\lozenge \winningstates | \initialstate) = 1\). Hence, \((\policy^{\playerone}, \policy^{\playertwo})\) is an equilibrium policy pair for every \(\policy^{\playerone} \in \Policy^{1, win}\).
\end{proof}

\begin{rem}
For clarity of presentation, we restrict the feasible policy spaces of the players to \(\Policy^{\playerone, St, win}\) and \(\Policy^{\playertwo, St}\). The equilibrium pair \((\policy^{\playerone,*}, \policy^{\playertwo,*})\in \Policy^{\playerone, St, win} \times \Policy^{\playertwo, St}\) of policies from Proposition \ref{prop:eqpolicies} also satisfy \[\sup_{\policy^{2} \in \Policy^{\playertwo}} C(\policy^{1}, \policy^{2,*})\leq C(\policy^{1,*}, \policy^{2,*}) \leq \inf_{\policy^{1} \in \Policy^{\playerone, win}} C(\policy^{1}, \policy^{2,*}).\]
\end{rem}

Knowing that the stationary policies suffice to find an equilibrium policy pair, we can represent the payoff of each step as a function of Player 1's policy and find a set of equilibrium policies via value iteration. Let \(\policy^{\playerone}(\genericstate)\) and \(\policy^{\average}(\genericstate)\) denote the vector of Player \(\playerone\)'s and average player's policies at state \(\genericstate\), respectively. Also, let \(v(\genericstate)\) denote the payoff-to-go at state \(\genericstate\) such that \(v(\genericstate) = 0\) for all \(\genericstate \in \winningstates\) and \(v(\genericstate) = \infty\) for all \(\genericstate \in \states \setminus \potentiallywinningstates\). By the first-order optimality conditions, for all \(\genericstate \in \states \setminus \potentiallywinningstates\), we have
\begin{linenomath*}
 \begin{equation*}
\begin{split}
\label{eqn:opteqn}
&v(\genericstate) = \min_{\policy^{\playerone}(\genericstate)} \Bigg( KL\left( \policy^{\playerone}(\genericstate)|| \policy^{\average}(\genericstate) \right)  +  \\
& \max_{\policy^{\playertwo}(\genericstate)} \sum_{\altstate \in \states } \sum_{\substack{ \genericaction^{\playerone} \in \actions^{1}(\genericstate)\\ \genericaction^{\playertwo} \in \actions^{\playertwo}(\genericstate)}} \probs(\genericstate,\genericaction^{\playerone}, \genericaction^{\playertwo},\altstate) \policy^{\playerone}(\genericstate,\genericaction^{\playerone}) \policy^{\playertwo}(\genericstate,\genericaction^{\playertwo}) v(\altstate) \Bigg)
\end{split} 
\end{equation*} 
\end{linenomath*} where the arguments of the minimum are the equilibrium policies for Player \(\playerone\). 
Similarly, by the first-order optimality conditions, we can show that for all \(\genericstate \in \states \setminus \potentiallywinningstates\), the equilibrium policies for Player \(\playertwo\) satisfy
\begin{linenomath*}
 \begin{equation*}
\begin{split}
\label{eqn:opteqn2}
&\policy^{\playertwo}(\genericstate) = \arg \max_{\policy^{\playertwo'}(\genericstate)} \sum_{\genericaction^{\playerone} \in \actions^{1}(\genericstate)}\policy^{\average}(\genericstate,\genericaction^{\playerone}) \\  &\exp\Bigg( \sum_{\substack{\altstate \in \states \\ \genericaction^{\playertwo} \in \actions^{\playertwo}(\genericstate)}} \probs(\genericstate,\genericaction^{\playerone}, \genericaction^{\playertwo},\altstate) \policy^{\playertwo'}(\genericstate,\genericaction^{\playertwo}) v(\altstate) \Bigg)^{-1}.
\end{split} 
\end{equation*} 
\end{linenomath*}

\section{Offline Learning of Player \(\playertwo\)'s Policy} \label{section:learning}
In this section, we give an algorithm to learn Player \(\playertwo\)'s policy and synthesize a near-optimal policy for Player \(1\). 

\textcolor{mycolor2}{Algorithm \ref{algo:learning} takes the game model \(\identityconcealmentgame\) and \(\samplepaths\) sample runs (with infinite lengths\footnote{\textcolor{mycolor2}{Since Algorithm \ref{algo:learning} utilizes only \(\sampleperstate\) samples per state, in practice, we need to store at most \(\sampleperstate \cardstates\) transitions.}}) collected under the average player's policy (Line \ref{algstep:collect}). A potentially winning state is \textit{known} if there are a total of at least \(\sampleperstate\) sample transitions from that state in the sample runs (Line \ref{algstep:known}). Otherwise, the state is \textit{unknown}. Let \((i,j)\) denote the label of the transition in the \(i\)-th sample run at time \(j\). For every known state \(\genericstate\), we create an ascending order of the sample transitions from \(\genericstate\) where the index of the sample runs has a higher priority. An example ordering is \((1,0),(1,3),(2,1)\). For every known state, Algorithm \ref{algo:learning} estimates Player \(\playertwo\)'s policy using the first \(\sampleperstate\) action samples from that state as \(\policy^{\playertwo}(\genericstate)\)(Line \ref{algstep:estimate}).
In Algorithm \ref{algo:learning}, we consider a modified game \(\identityconcealmentgame'\) (Lines \ref{algstep:setdefinitions}-\ref{algstep:modifiedgame}) where the unknown states are also in the winning states. After constructing \(\identityconcealmentgame'\), we solve for the optimal winning policy \(\policy^{\playerone,'}\) when Player \(\playertwo\)'s policy is the estimated policy \(\policy^{\playertwo}\) (Line \ref{algstep:optimalknown}). The output policy \(\policy^{\playerone}\) for the original game \(\identityconcealmentgame\) is history-dependent and uses one-bit of extra memory compared to a stationary policy. The memory bit tracks whether an unknown state has been visited yet. The output policy \(\policy^{\playerone}\) uses the optimal policy \(\policy^{\playerone,'}\) (synthesized in Line \ref{algstep:optimalknown}) against Player \(\playertwo\)'s estimated policy until reaching an unknown state. If an unknown state has been visited in the history, the output policy uses the average player's policy. 
}

\textcolor{mycolor2}{
Algorithm 1 considers a modified game where the unknown states are also in the winning states. This game construction ensures that the optimal value for the modified game is lower than that of the original game when the respective sets of winning policies are considered. After reaching an unknown state, Player \(\playerone\) follows the average player's policy and induces zero KL divergence payoff. The policy construction ensures that Player \(1\)'s policy has the same objective value for both the modified and the original games and, consequently, is near-optimal for the original game. On the other hand, after reaching an unknown state, Player 1 may not win the original game since it does not necessarily take safe actions. While the learned policy may not be a winning policy, we later show that reaching an unknown state and losing the game happens with a low probability. 
}

	\begin{algorithm} [t]
	\caption{Offline learning of Player \(\playertwo\)'s policy and policy optimization for Player \(\playerone\)} \label{algo:learning}
	\begin{algorithmic}[1]
		\State \textbf{Input:} An identity concealment game \(\identityconcealmentgame\), \(\samplepaths\) independent sample runs under (\(\policy^{\average}\),\(\policy^{\playertwo, \circ}\)). \label{algstep:collect}
		\State \textbf{Output:} A policy \(\policy^{\playerone}\) for Player \(\playerone\).
		\State \(\knownstates := \emptyset\).
		\For{\(\genericstate \in \potentiallywinningstates\)}
		\If{\(\hat{\sampleperstate}_{\genericstate} \geq \sampleperstate \)} 
  		\State \(\knownstates := \knownstates \cup \lbrace \genericstate \rbrace\) \label{algstep:known}.
		\State Set \(\policy^{\playertwo}(\genericstate)\) as the empirical distribution of first \(\sampleperstate\) actions of Player \(\playertwo\) at state \(\genericstate\). \label{algstep:estimate}
		\EndIf
		\EndFor 
		\State \(\unknownstates := \potentiallywinningstates \setminus \knownstates\), \(\statesend := \unknownstates \cup \winningstates\). \label{algstep:setdefinitions}
		\State Construct a modified identity concealment game \(\identityconcealmentgame'\) that is the same as \(\identityconcealmentgame\) except that all states in \(\statesend\) are absorbing, and \(\statesend\) is the set of winning states. \label{algstep:modifiedgame}
		\State For \(\identityconcealmentgame'\), synthesize the optimal stationary policy \(\policy^{\playerone,'}\) using the estimated policy \(\policy^{\playertwo}(\genericstate)\) for all \(\genericstate \in \knownstates\). \label{algstep:optimalknown}
  \textcolor{mycolor2}{
  \For{\(t = 0, \ldots \)}
  \State For every \(h_{t} = s_{0}a^{\playerone}_{0}a^{\playertwo}_{0}\ldots  s_{t}\), define \(\mu_{t}^{\playerone}(h_{t})\) such that \(\mu_{t}^{\playerone}(h_{t}):= \policy^{\playerone,'}(\genericstate_{t})\) if \(\genericstate_{i} \in \knownstates\) for all \(0 \leq i \leq t\),  and  \(\mu_{t}^{\playerone}(h_{t}) := \policy^{\average}(\genericstate_{t})\) otherwise. \label{algstep:unknownaverage}
  \EndFor
		\State For \(\identityconcealmentgame\), define the policy \(\policy^{\playerone} := \mu_{0}^{\playerone} \mu_{1}^{\playerone} \ldots\). \label{algstep:unknownaverage}}
	\end{algorithmic}
\end{algorithm}

We define some notation before discussing the properties of the algorithm. The equilibrium value \( C(\policy^{1,*}, \policy^{2,*})\) of the game is denoted by \(\gamevalue^{*}\). Note that there is a unique value \(\gamevalue^{*}\) due to Proposition \ref{prop:eqpolicies}. Player \(\playertwo\)'s true policy is \(\policy^{ \playertwo,\circ}\). For a state \(\genericstate\), the total number of collected sample transitions from \(\genericstate\) is \(\hat{\sampleperstate}_{\genericstate}\), and the empirical action frequencies for Player \(\playertwo\) using only the first \(\sampleperstate\) samples drawn from \(\policy^{\playertwo,\circ}\) is \(\policy^{\playertwo}\). Player \(1\)'s optimal winning policy against \(\policy^{\playertwo,\circ}\) is \(\policy^{\playerone,\circ}\), i.e., \(\policy^{\playerone,\circ} = \arg\min_{\policy^{\playerone} \in \Policy^{1, win}} C(\policy^{1}, \policy^{2,\circ}) =\arg\min_{\policy^{\playerone} \in \Policy^{1, win}}  \mathbb{E}\left[\sum_{t=0}^{\randomstoppingtime} KL(\mu^{\playerone}_{t}(\history_{t})|| \policy^{\average}(\genericstate_{t})) | \policy^{\playerone}, \policy^{\playertwo, \circ} \right]\). For \(\potentiallywinningstates\), \(\maxcost\) denotes the maximum KL divergence between the safe action distributions for Player \(\playerone\) and the action distribution for \(\policy^{\average}\), i.e., \(\maxcost = \max_{\genericstate \in \potentiallywinningstates, \genericaction^{\playerone} \in \actions^{\playerone}} (\log(\policy^{\average}(\genericstate))^{-1}\) subject to \(\genericaction^{1}\) is safe.

We have the following assumption on Player \(\playertwo\)'s policy. Assumption \ref{ass:stationarityofpursuer} ensures tractability of estimation for the transition probabilities.
\begin{assum} \label{ass:stationarityofpursuer}
	\(\policy^{\playertwo,\circ}\) is stationary on \(\states\).
\end{assum}

Algorithm \ref{algo:learning} satisfies the requirements given in Problem \ref{problem:learning} in two steps: \(1)\) The objective value incurred by \(\policy^{\playerone}\) is close to the optimal value for the known states since Player \(\playertwo\)'s policy will be estimated accurately for these states. For the unknown states, \(\policy^{\playerone}\) will incur \(0\) payoff since \(\policy^{\playerone}\) is the same with \(\policy^{\average}\) for these states. Overall, the KL objective value will be close to the optimal value under \(\policy^{\playerone}\). \(2)\) If the number of sample runs is large enough, unknown states are reached with low probability under \(\policy^{\average}\). If the unknown states are visited with high probability under \(\policy^{\playerone}\), then the objective value would be large since the deviation of \(\policy^{\playerone}\) from \(\policy^{\average}\) would be large. However, since the KL objective function is near optimal, the unknown states are visited with low probability under \(\policy^{\playerone}\), and the probability of losing is low for Player \(\playerone\).  

 We define that a stationary policy pair \((\policy^{\playerone}, \policy^{\playertwo})\) has an \textit{\((L,\discount')\)-contraction}, if  \(\min_{\genericstate \in  \potentiallywinningstates} \Pr\left(\lozenge_{\leq L} \winningstates \cup \unknownstates |\initialstate = \genericstate\right) \) \(\geq\) \(1 - \discount'.\) To show the near optimality of the output policy, we make the following assumption, which ensures the finiteness of the expected length of a game run.

\begin{assum} \label{ass:contractionassumption}
	The policy pair \((\policy^{\playerone}, \policy^{\playertwo})\) has an \(\left(L, \discount - \frac{\optimalitygap (1-\discount)^2}{\maxcost L}\right)\)-contraction where \(\discount\) is a constant strictly lower than \(1\).
\end{assum}
The validity of Assumption \ref{ass:contractionassumption} can be checked after the termination of the algorithm since both \(\policy^{\playerone}\) and \( \policy^{\playertwo}\) are known.  If the assumption is violated, one can increase \(\discount\) and rerun the algorithm. We remark that \(\discount - (\optimalitygap (1-\discount)^2)/(\maxcost L)\) is an increasing function of \(\discount\), and the policy pair \((\policy^{\playerone}, \policy^{\playertwo})\) has to have a \((\cardstates, \discount - (\optimalitygap (1-\discount)^2)/(\maxcost \cardstates))\)-contraction for some \(\discount < 1\) since otherwise \(\policy^{\playerone}\) has to incur infinite payoff.  Therefore, there always exists \(\discount < 1\) that satisfies the assumption. We note that having a contraction, e.g., a discount factor, is a common assumption in reinforcement learning to ensure the boundedness of the objective function~\citep{sutton2018reinforcement}.

The following proposition shows that Algorithm \ref{algo:learning} indeed results in a near-optimal policy using only the game runs collected under the average player's policy.
\begin{prop} \label{proposition:learning}
	Let \( w  =  (  \gamevalue^{*} +  \log(2) + \optimalitygap)/\losingprobability\). Under Assumptions \ref{ass:stationarityofaverage}, \ref{ass:stationarityofpursuer}, and \ref{ass:contractionassumption}, if \[\sampleperstate \geq \frac{4 \maxcost^{2} L^{4}\left(2 \log(2) \cardactions + \log\left({2\cardstates}/{\failureprob} \right) \right) }{ (1-\discount)^{4} \optimalitygap^{2}  }\] and \[ \samplepaths \geq {e^{2w}}  \log\left( 4/\failureprob \right)/2 + 2 \cardstates e^{w}\sampleperstate\] in Algorithm \ref{algo:learning}, then the output policy \(\policy^{\playerone}\) satisfies \[C(\policy^{1}, \policy^{2,\circ}) -  C(\policy^{1, \circ}, \policy^{2,\circ}) \leq \optimalitygap\] and \[{\Pr}^{\policy^{\playerone}, \policy^{\playertwo,\circ}}(\lozenge \winningstates | \initialstate) \geq 1- \losingprobability,\] with probability at least \(1 - \failureprob\).
\end{prop}

We use a series of lemmas to prove Proposition \ref{proposition:learning}. Lemma \ref{lemma:empricalisclose} shows that with high probability, the estimated action distribution \(\policy^{\playertwo}\) and the actual action distribution \(\policy^{\playertwo, \circ}\) are close for all known states\footnote{\textcolor{mycolor2}{One can use all available sample transitions instead of the first \(m\) samples. While this approach yields a concentration bound of the same order (see Lemma 3 of \citep{karabag2023sample}) and may improve the estimates for some states, we use only the first  \(m\) samples for every state since the performance bound given in Lemma \ref{lemma:closemdpcloseoutcome} requires uniform coverage.}}. The proof follows Sanov's theorem and Pinsker's inequality combined with the union bound~\citep{weissman2003inequalities}.
\begin{lem}[] \label{lemma:empricalisclose}
 For any \(\failureprob_{\known} \in (0,1]\),  given \(m\) independent transitions sampled from \(\policy^{\playertwo,\circ}(\genericstate)\) for each \(\genericstate \in \knownstates\), with probability at least \(1 - \failureprob_{\known}\), \[ \left \Vert \policy^{\playertwo,\circ}(\genericstate) - \policy^{\playertwo}(\genericstate) \right \Vert_{1} \leq \sqrt{\frac{2(\log(2) \cardactions + \log(\cardstates/\failureprob_{\known}))}{\sampleperstate}}\]  for all \(\genericstate \in \knownstates\).
\end{lem}

Lemma \ref{lemma:closemdpcloseoutcome} shows that if the estimated and actual transition probability distributions are close, and the policy pair \((\policy^{\playerone}, \policy^{\playertwo,\circ})\) has an \(L\)-step contraction, then the values of the objective function are close for \((\policy^{\playerone}, \policy^{\playertwo,\circ})\) and \((\policy^{\playerone}, \policy^{\playertwo})\). The paper \citep{strehl2008analysis} showed this property for \((1, \discount)\)-contractions. We extend this result to \((L, \discount)\)-contractions and show that difference between the value functions is bounded by representing the KL objective as a sum of payoffs per time step. Since \((\policy^{\playerone}, \policy^{\playertwo})\) has \(L\)-step contraction lower than or equal to \(\discount - {\optimalitygap (1-\discount)^2}/{(\maxcost L)}\) and \(\Vert \policy^{\playertwo,\circ}(\genericstate) - \policy^{\playertwo}(\genericstate) \Vert_{1} \leq {\optimalitygap (1-\discount)^2}/{(\maxcost L^{2})}\) for all \(\genericstate \in \knownstates\), then \((\policy^{\playerone}, \policy^{\playertwo, \circ})\) has \((L, \discount )\)-contraction. Since \((\policy^{\playerone}, \policy^{\playertwo, \circ})\) has \((L, \discount )\)-contraction, the KL objective value is bounded by \(\sum_{i=0}^{\infty} L \maxcost \discount^{i} = L \maxcost/(1-\discount)\) from every initial state in \(\potentiallywinningstates\setminus\winningstates\). Because the estimated and true transition probabilities are close as in Lemma \ref{lemma:closemdpcloseoutcome}, and \((\policy^{\playerone}, \policy^{\playertwo, \circ})\) has \((L, \discount )\)-contraction, the \( \|\Gamma^{\policy^{\playerone}, \policy^{\playertwo}} - \Gamma^{\policy^{\playerone}, \policy^{\playertwo,\circ}} \|_{1}\) is bounded by \(\optimalitygap(1-\discount)/(2 L \maxcost)\). Since the KL objective value is bounded from every initial state and the distributions of game runs induced by \(({\policy^{\playerone}, \policy^{\playertwo}})\) and \(({\policy^{\playerone}, \policy^{\playertwo,\circ}})\) are close to each other, the KL objective differs by at most \(\optimalitygap/2\). 

\begin{lem}[] \label{lemma:closemdpcloseoutcome}
 If \(\Vert \policy^{\playertwo,\circ}(\genericstate) - \policy^{\playertwo}(\genericstate) \Vert_{1} \leq \frac{\optimalitygap (1-\discount)^2}{\maxcost L^{2}}\) for all \(\genericstate \in \knownstates\), \[\vert C(\policy^{1}, \policy^{2})-C(\policy^{1}, \policy^{2,\circ}) \vert  \leq \frac{\optimalitygap}{2}.\]
\end{lem}

The following lemmas show that the probability of losing is low if the number of sample trajectories is high. Lemma \ref{lemma:lowsamplelowreachChernoffImproved} shows that if a state is unknown, then the probability of reaching that state is low. The proof is an application of the Chernoff-Hoeffding inequality. We use the fact that number of collected action samples from a state is higher than the number of sample runs that visit the state. Since the unknown states does not have enough sample transitions, it implies that these states are visited with a low probability.

\begin{lem}[] \label{lemma:lowsamplelowreachChernoffImproved}
	 Let \(\hat{\sampleperstate}_{\genericset}\) denote the number of transitions from set \(\genericset\) of states using \(\samplepaths\) runs independently sampled under policies \((\policy^{\average},\policy^{\playertwo,\circ})\). For  \(\sampleperstate' \geq \hat{\sampleperstate}_{\genericset}\) and \(1/2 \geq \sigma > \sampleperstate'/\samplepaths\), with probability at least \(1 - 2 \exp(-2n (\sigma - \sampleperstate'/\samplepaths)^{2})\), we have \({\Pr}^{\policy^{\average}, \policy^{\playertwo,\circ}}(\lozenge \genericset |\initialstate) \leq \sigma\).
\end{lem}

Lemma \ref{lemma:highreachhighcostImproved} shows that if a state is reached with high probability under \(\policy^{\playerone}\) and with low probability under \(\policy^{\average}\), then the value of the objective function is high. The proof follows from that the KL divergence between the distributions of game runs is lower bounded by the KL divergence between the reachability probability to set \(\genericset\) by the data processing inequality. Since the unknown states are visited with low probability under \(\policy^{\average}\), visiting these states with high probability causes deviations from the average player and incurs a high value for the KL objective function.

\begin{lem}[] \label{lemma:highreachhighcostImproved}
Let \(\genericset \subseteq \states\). If \({\Pr}^{\policy^{\playerone}, \policy^{\playertwo,\circ}}(\lozenge \genericset |\initialstate) > \frac{ -(\gamevalue^{*} + \log(2) + \optimalitygap)} {\log\left({\Pr}^{\policy^{\average}, \policy^{\playertwo,\circ}}(\lozenge \genericset |\initialstate)\right)}\), then \(C(\policy^{1}, \policy^{2,\circ}) > \gamevalue^{*} + \optimalitygap.\)
\end{lem}

The proof of Proposition \ref{proposition:learning} consists of two parts. We first show that the learned policy \(\policy^{\playerone}\) is near optimal. At the unknown states, \(\policy^{\playerone}\) is the same with the average player's policy \(\policy^{\average}\)  and incurs \(0\) payoff. Consider a modified identity concealment game where the unknown states are included in the winning states. The equilibrium value of the modified game is less compared to the original game. For the known states, Player \(\playertwo\)'s estimated policy will be close to the true policy due to Lemma \ref{lemma:empricalisclose}. Since the estimated policy is accurate, \(\policy^{\playerone}\) is near optimal for the modified game. Thus, \(\policy^{\playerone}\) is near optimal due to Lemma \ref{lemma:closemdpcloseoutcome}. To show that the probability of losing is small, we use Lemmas \ref{lemma:lowsamplelowreachChernoffImproved} and \ref{lemma:highreachhighcostImproved}. Since the ratio between the numbers of sample paths and sample transitions is high enough, Lemma \ref{lemma:lowsamplelowreachChernoffImproved} implies that the probability of reaching an unknown state is bounded under \(\policy^{\average}\). Since \(\policy^{\playerone}\) is near optimal, the probability of reaching an unknown state is also bounded under the learned policy due to Lemma \ref{lemma:highreachhighcostImproved}.

\begin{proof}[Proof of Proposition \ref{proposition:learning}]

	At time \(t\) define \(\policy^{\playerone, \triangleleft}\) such that  \(\policy^{\playerone, \triangleleft}(\genericstate)\) \( := \policy^{\playerone , \circ}(\genericstate)\) if \(\genericstate_{i} \in \knownstates\) for all \(0 \leq i < t\),  and  \(\policy^{\playerone, \triangleleft}(\genericstate) := \policy^{\average}(\genericstate)\) otherwise. 
	For notational convenience, define \(w := {(  \gamevalue^{*} +  \log(2)  + \optimalitygap)}/{\losingprobability}\).

	By Lemma \ref{lemma:empricalisclose}, if \(\sampleperstate \geq \frac{4 \maxcost^{2} L^{4}\left(2 \log(2) |\states^{\playertwo}|+ \log\left(\frac{2}{\failureprob} \right) \right)}{(1-\discount)^{4} \optimalitygap^{2} }\) in Algorithm \(1\), then with probability at least \(1 -\failureprob/2\), we have \(\left \Vert \policy^{\playertwo,\circ}(\genericstate) - \policy^{\playertwo}(\genericstate) \right \Vert_{1}  \leq {\optimalitygap (1-\discount)^2}/{(\maxcost L^{2})}\) for all \(\genericstate \in \knownstates\)\footnote{\textcolor{mycolor2}{The concentration bound given in in the lemma requires independent transitions. However, the transition in the sample runs may not be independent in general. Despite the dependent samples, the concentration bound can still be used for our analysis. A detailed discussion on the dependence of sample transitions and the use of this bound is given in \citep{strehl2008analysis}. Alternatively, one can use a concentration bound that can handle dependent transitions and random stopping times (e.g., Lemma 3 of \citep{karabag2023sample}) and get the same order of convergence.}}. Then by Lemma \ref{lemma:closemdpcloseoutcome}, we have \(|C(\policy^{1,\triangleleft}, \policy^{2,\circ}) - C(\policy^{1,\triangleleft}, \policy^{2})| \leq \optimalitygap/2\) and \(|C(\policy^{1}, \policy^{2}) - C(\policy^{1}, \policy^{2,\circ})| \leq \optimalitygap/2\) with probability at least \(1- \failureprob/2\). Since \(C(\policy^{1}, \policy^{2}) \leq C(\policy^{1,\triangleleft}, \policy^{2})\) due to the optimality of \(\policy^{\playerone}\) against \(\policy^{\playertwo}\), we have \(|C(\policy^{1,\triangleleft}, \policy^{2,\circ}) - C(\policy^{1}, \policy^{2,\circ})| \leq \optimalitygap\) with probability at least \(1- \failureprob/2\). We also have that \(C(\policy^{1,\triangleleft}, \policy^{2,\circ}) \leq C(\policy^{1,\circ}, \policy^{2,\circ})\) since \(\policy^{\playerone , \circ}\) and \(\policy^{\playerone, \triangleleft}\) induce the same payoff until reaching an unknown state, and \(\policy^{\playerone , \circ}\) induces a non-negative payoff after reaching an unknown state whereas \(\policy^{\playerone, \triangleleft}\) induces \(0\) payoff after reaching an unknown state. Consequently, with probability at least \(1- \failureprob/2\), we have \[ C(\policy^{1}, \policy^{2,\circ}) \leq  C(\policy^{1,\circ}, \policy^{2,\circ}) + \optimalitygap.\] 
	
	We now show that if \(\samplepaths/\sampleperstate \geq  e^{2w} \log\left( 4/\failureprob \right)/ 2m + 2 \cardstates e^{w}\) and \(\sampleperstate\) is as above, \({\Pr}^{\policy^{\playerone}, \policy^{\playertwo}}(\lozenge \winningstates | \initialstate) \geq 1- \losingprobability\) with probability at least \(1 - \failureprob/2\). Let \(\unknownstates\) be the set of unknown states. Define \(y :=  {e^{w} \log\left( {4}/{\failureprob} \right) }/{(2m \cardstates)}\) and \(c :=  y + 2 \). We have  \({\samplepaths}/{\sampleperstate} \geq  \cardstates c e^{w} .\) Also define \(c': = \left( y + \sqrt{y} \sqrt{y+4} + 2 \right)/2.\) Note that \(y \geq 0\) and \(c \geq c' \geq 1\). The number of sample transitions from \(\unknownstates\) is lower than \(\sampleperstate\cardstates\) by the definition. By Lemma 6, with probability at least \(1-2\exp\left( -2 \samplepaths (1-1/c)^{2}  e^{-2w}  \right)\), we have \({\Pr}^{\policy^{\average}, \policy^{\playertwo, \circ}}(\lozenge \unknownstates | \initialstate) \) \(\leq\) \(e^{-w}\). Since \(c \geq c' \geq 1\), we have 
		\(2\exp  \left( -2 \samplepaths \right(\frac{c-1}{c}\left)^{2}  e^{-2w}  \right) \allowbreak \leq \allowbreak	2\exp\left( - 2m\cardstates c'\right(\frac{c'-1}{c'}\left)^{2}     e^{-w}  \right) \allowbreak = \allowbreak \failureprob/2\). Thus, we have \({\Pr}^{\policy^{\average}, \policy^{\playertwo, \circ}}(\lozenge \unknownstates | \initialstate)\leq e^{-w}\) with probability at least \(1-\failureprob/2\). 
	
	If \(  C(\policy^{1}, \policy^{2,\circ}) \leq  C(\policy^{1,\circ}, \policy^{2,\circ}) + \optimalitygap\), we have \(C(\policy^{1}, \policy^{2,\circ}) \leq \gamevalue^{*} + \optimalitygap\) since \(C(\policy^{1,\circ}, \policy^{2,\circ}) \leq \gamevalue^{*}.\) By Lemma 7, the probability \({\Pr}^{\policy^{\playerone}, \policy^{\playertwo, \circ}}(\lozenge \unknownstates | \initialstate)\)  \(\leq\) \(\losingprobability\) with probability at least \(1-\failureprob\) since \({\Pr}^{\policy^{\average}, \policy^{\playertwo, \circ}}(\lozenge \unknownstates | \initialstate)\)  \(\leq\) \(e^{-w}\) with probability at least \(1-\failureprob/2\). Since Player \(\playerone\) can lose the game only by reaching an unknown state, the probability of losing is at most \(\losingprobability\)  with probability at least \(1-\failureprob/2\).
	
	Combining the near-optimality result for the objective function and the result for the probability of losing, we conclude that \(\policy^{\playerone}\) satisfies \[ C(\policy^{1}, \policy^{2,\circ}) \leq  C(\policy^{1,\circ}, \policy^{2,\circ}) + \optimalitygap.\] and \[{\Pr}^{\policy^{\playerone}, \policy^{\playertwo}}(\lozenge \winningstates | \initialstate) \geq 1- \losingprobability\] with probability at least \(1 - \failureprob\). 
\end{proof}

\section{Numerical Examples}
In this section, we give numerical examples of the equilibrium policies for identity concealment games and offline policy optimization for Player \(\playerone\). 

\subsection{Detection of Hostile Clients in Cyber Interactions}

\begin{figure}[t]
		\centering
        \input{htesting}
	\centering
	\caption{Receiver operating characteristic curve of the likelihood ratio classifier that identifies hostile clients. True positive rate is the ratio of detected attackers to all attackers. False positive rate is the ratio of real clients identified as an attacker to all clients.} 
	\label{fig:htesting}
\end{figure}

\label{section:detectionexample}
We show the effect of identity concealment on the detection of hostile clients in the cyber interaction scenario shown in Figure \ref{fig:dosattack}. The game is played between a client (Player \(\playerone\)) and the server (Player \(\playertwo\)), and the states represent the remaining times for the client's processed requests, if there are any. At every time, the client can disconnect, make a request, or wait. The server can accept or reject the client's potential request. However, the server cannot reject the request again if the client has been rejected previously for that request. If the request is accepted, it takes a certain number of steps to process the request. The attacker's goal is to cause a denial of service by overwhelming the server, and it wins the game if and only if the server concurrently processes multiple requests of the client. At every state, the real clients' policy, i.e, the average player's policy, is randomized, and is more likely to make a request if there are no requests being processed or there is a rejected request. The details of the setting are given in \citep{karabag2021identity}.

 In Figure \ref{fig:htesting}, we observe that when the server uses its equilibrium policy \(\policy^{\playertwo, *}\), hostile clients are identified with high accuracy compared to policy \(\policy^{\playertwo, U}\) that accepts or rejects the requests with equal probabilities. This is because, unlike \(\policy^{\playertwo, U}\), the equilibrium policy \(\policy^{\playertwo, *}\) is state-dependent, and using \(\policy^{\playertwo, *}\) the server can drive the game into a state where the hostile client's behavior is different from the real clients' behaviors. Similarly, a hostile client is less likely to be detected when it uses its equilibrium policy \(\policy^{\playerone, *}\) compared to the greedy policy \(\policy^{\playerone, G}\) that makes a request at every time step. We also observe that an additional interaction, i.e., a game run, improves the accuracy of classification as explained in Section \ref{section:problemstatement}.

	\subsection{Equilibrium Policies for a Pursuit-Evasion Game}

	We show the behavior for hostile Player \(\playerone\) in a pursuit-evasion game. Player \(\playerone\) is an evader and Player \(\playertwo\) is a pursuer. The environment is a two-dimensional grid where each node represents an intersection. At each time step, every intersection is occupied with probability \(0.5\). If the pursuer's intersection is clear, it can move in \(+x\), \(-x\), \(+y\), \(-y\) directions by \(1\) or stay at its current intersection. If the intersection is occupied, the pursuer stays. Regardless of the state of its intersection, the evader can move in all directions by \(1\) or \(2\) blocks, or stay at the current intersection. We encode the states of the game using the relative distances on \(x\) and \(y\) coordinates between the players, the states of the evader's intersection, and the state of the pursuer's intersection. If the block distance between the players is \(6\) or more the evader wins the game, and if the block distance is 0, the pursuer wins the game, i.e., traps the average evader.

The average player's policy only depends on whether its intersection is occupied and is given in Table \ref{table:eqpolicies}. If the intersection is occupied, with high probability the average player waits the intersection to be cleared and with  low probability the average player ``illegally'' crosses the intersection. If its intersection is clear, with high probability the average player moves with a normal speed, otherwise it ``overspeeds'', i.e., moves by \(2\) blocks.

The equilibrium policy of the hostile evader is given in Table \ref{table:eqpolicies} for some states. If the evader is close to the pursuer, for instance at state \((1,0,C,C)\), the evader tends behave more suspiciously by ``overspeeding''. When the distance is higher, for instance at state \((2,0,C,C)\), the evader does not flee and behaves more closely to the average player. Similarly, the hostile evader chooses to move at state \((1,0,O,C)\) since otherwise the evader may get caught. On the other hand, the hostile evader chooses to stay  at state \((1,0,O,O)\) with high probability since the pursuer has to wait. Overall, the hostile evader behaves dissimilar to the average player only if it is in a dangerous position, which matches the intuition of concealing the identity by behaving in an unsuspicious way.

\begin{table}[t]
\caption{The policies of the average and hostile evaders. O denotes an occupied intersection and C denotes a clear intersection.}
 \label{table:eqpolicies} 
\resizebox{\columnwidth}{!}{%
\begin{tabular}{cc|ccccccccc|}
\cline{3-11}
\multicolumn{1}{l}{}                                                          & \multicolumn{1}{l|}{}                                                                & \multicolumn{9}{c|}{Probability of Action}                                                                                                                                                                                                                                     \\ \hline
\multicolumn{1}{|c|}{\begin{tabular}[c]{@{}c@{}}Evader\\ Type\end{tabular}} & State                                                                                & Stay                         & x                            & 2x                           & y                            & 2y                           & -y                           & -2y                          & -x                        & -2x                       \\ \hline
\multicolumn{1}{|c|}{}                                                        & \begin{tabular}[c]{@{}c@{}}Intersection\\ Occupied\end{tabular}                      & 0.80                         & 0.10                         & 0                            & 0.05                         & 0                            & 0.05                         & 0                            & 0                         & 0                         \\
\multicolumn{1}{|c|}{\multirow{-2}{*}{Average}}                               & \cellcolor[HTML]{EFEFEF}\begin{tabular}[c]{@{}c@{}}Intersection\\ Clear\end{tabular} & \cellcolor[HTML]{EFEFEF}0    & \cellcolor[HTML]{EFEFEF}0.40 & \cellcolor[HTML]{EFEFEF}0.10 & \cellcolor[HTML]{EFEFEF}0.20 & \cellcolor[HTML]{EFEFEF}0.05 & \cellcolor[HTML]{EFEFEF}0.20 & \cellcolor[HTML]{EFEFEF}0.05 & \cellcolor[HTML]{EFEFEF}0 & \cellcolor[HTML]{EFEFEF}0 \\ \hline
\multicolumn{1}{|c|}{}                                                        & (2,0,\(C\),\(C\))                                                                            & 0                            & 0.41                         & 0.23                         & 0.12                         & 0.06                         & 0.12                         & 0.06                         & 0                         & 0                         \\
\multicolumn{1}{|c|}{}                                                        & \cellcolor[HTML]{EFEFEF}(1,0,\(C\),\(C\))                                                    & \cellcolor[HTML]{EFEFEF}0    & \cellcolor[HTML]{EFEFEF}0.31 & \cellcolor[HTML]{EFEFEF}0.33 & \cellcolor[HTML]{EFEFEF}0.09 & \cellcolor[HTML]{EFEFEF}0.09 & \cellcolor[HTML]{EFEFEF}0.09 & \cellcolor[HTML]{EFEFEF}0.09 & \cellcolor[HTML]{EFEFEF}0 & \cellcolor[HTML]{EFEFEF}0 \\
\multicolumn{1}{|c|}{}                                                        & (1,0,\(O\),\(C\))                                                                            & 0                            & 0.62                         & 0                            & 0.19                         & 0                            & 0.19                         & 0                            & 0                         & 0                         \\
\multicolumn{1}{|c|}{\multirow{-4}{*}{Hostile}}                               & \cellcolor[HTML]{EFEFEF}(1,0,\(O\),\(O\))                                                    & \cellcolor[HTML]{EFEFEF}0.74 & \cellcolor[HTML]{EFEFEF}0.16 & \cellcolor[HTML]{EFEFEF}0    & \cellcolor[HTML]{EFEFEF}0.05 & \cellcolor[HTML]{EFEFEF}0    & \cellcolor[HTML]{EFEFEF}0.05 & \cellcolor[HTML]{EFEFEF}0    & \cellcolor[HTML]{EFEFEF}0 & \cellcolor[HTML]{EFEFEF}0 \\ \hline
\end{tabular}%
}
\end{table}

\subsection{Offline Learning of the Pursuer's Policy}
In this example, we show the empirical performance of the proposed offline learning algorithm for different number of sample runs \(\samplepaths\) and number of estimation samples \(\sampleperstate\) per state. Note that we do not give optimality guarantees for the demonstrated values of \(\sampleperstate\) and \(\samplepaths\). We use the same environment with the previous example where the initial state \(\initialstate\) is \((1,0,O,O.)\). The pursuer's policy \(\policy^{\playertwo, \circ}\) is defined as follows. At each time step the pursuer stops tracking the evader, and the evader wins with probability \(0.2\). If the pursuer does not stop, it takes allowed actions with uniform probabilities.

\begin{figure}[t]
	\centering
	    \begin{subfigure}[b]{0.5\columnwidth}
		\centering
%
%
\definecolor{mycolor1}{rgb}{0.00000,0.44700,0.74100}%
\definecolor{mycolor2}{rgb}{0.85000,0.32500,0.09800}%
\definecolor{mycolor3}{rgb}{0.92900,0.69400,0.12500}%
\definecolor{mycolor4}{rgb}{0.49400,0.18400,0.55600}%
\definecolor{mycolor5}{rgb}{0.46600,0.67400,0.18800}%
\definecolor{mycolor6}{rgb}{0.30100,0.74500,0.93300}%
\begin{tikzpicture}[scale=0.5]

\begin{axis}[%
width=1.6\textwidth,
height=1.3\textwidth,
at={(0in,0in)},
scale only axis,
xmode=log,
xmin=30,
xmax=1000000,
xtick={10,100,1000,10000,100000,1000000},
xticklabels={{\(\text{10}^\text{1}\)},{\(\text{10}^\text{2}\)},{\(\text{10}^\text{3}\)},{\(\text{10}^\text{4}\)},{\(\text{10}^\text{5}\)},{\(\text{10}^\text{6}\)}},
xminorticks=true,
ytick={0,0.05,0.1, 0.15},
yticklabels={{\(\text{0}\)},{\(\text{0.05}\)},{\(\text{0.1}\)},{\(\text{0.15}\)}},
yminorticks=true,
xlabel style={font=\color{white!15!black}, font=\large},
xlabel={m},
ymin=0,
ymax=0.15,
ylabel style={font=\color{white!15!black},font=\large},
ylabel={\( C(\policy^{1}, \policy^{2,\circ})\)},
axis background/.style={fill=white},
axis x line*=bottom,
axis y line*=left,
xmajorgrids,
xminorgrids,
ymajorgrids,
legend style={at={(0.03,0.03)}, anchor=south west, legend cell align=left, align=left, draw=white!15!black}
]
\addplot [color=mycolor1, line width=2.0pt, mark=o, mark options={solid, mycolor1}]
  table[row sep=crcr]{%
30	0.12203082723725\\
100	0.12091714286142\\
300	0.11844687786604\\
1000	0.0812345055261562\\
3000	0.0570473224165736\\
10000	0.000149699019724941\\
30000	0.000146916893409124\\
100000	9.99579187289684e-05\\
300000	0\\
};
\addlegendentry{\(n=10^5\)}

\addplot [color=mycolor2, line width=2.0pt, mark=square*, mark options={solid, mycolor2}]
  table[row sep=crcr]{%
30	0.122388305491184\\
100	0.121753532005023\\
300	0.120850876567302\\
1000	0.118362858561549\\
3000	0.0818276689053114\\
10000	0.0187168901602229\\
30000	0.000149705213200534\\
100000	0.000146906377200036\\
300000	9.99584646043425e-05\\
};
\addlegendentry{\(n=3\times10^5\)}

\addplot [color=mycolor3, line width=2.0pt, mark=diamond*, mark options={solid, mycolor3}]
  table[row sep=crcr]{%
30	0.122355580017168\\
100	0.122267797146577\\
300	0.121838726556168\\
1000	0.120843165181019\\
3000	0.118438745891588\\
10000	0.0818229772474771\\
30000	0.0570458166238212\\
100000	0.000149720617261432\\
300000	0.000146905779999903\\
};
\addlegendentry{\(n=10^6\) }

\addplot [color=mycolor4, line width=2.0pt, mark=triangle*, mark options={solid, mycolor4}]
  table[row sep=crcr]{%
30	0.122663871449986\\
100	0.122245696628779\\
300	0.122165641983352\\
1000	0.121696897496646\\
3000	0.120834155074038\\
10000	0.118401501869921\\
30000	0.0818199463458179\\
100000	0.0187106464264871\\
300000	0.000149718412677376\\
};
\addlegendentry{\(n=3\times10^6\) }

\addplot [color=mycolor5, line width=2.0pt, mark=asterisk, mark options={solid, mycolor5}]
  table[row sep=crcr]{%
30	0.122338119736483\\
100	0.122284467589913\\
300	0.12228326072445\\
1000	0.122137775010795\\
3000	0.121788023314575\\
10000	0.120828094635158\\
30000	0.118429821785342\\
100000	0.0818200392529622\\
300000	0.0570453267170015\\
};
\addlegendentry{\(n=10^7\) }

\addplot [color=mycolor6, dashed, line width=2.0pt]
  table[row sep=crcr]{%
30	0.1222\\
100	0.1222\\
300	0.1222\\
1000	0.1222\\
3000	0.1222\\
10000	0.1222\\
30000	0.1222\\
100000	0.1222\\
300000	0.1222\\
};
\addlegendentry{Opt. Win}

\end{axis}

\end{tikzpicture}%
		\caption{}
		\label{fig:klobj}
	\end{subfigure}%
	\begin{subfigure}[b]{0.5\columnwidth}
		\centering
%
%
\definecolor{mycolor1}{rgb}{0.00000,0.44700,0.74100}%
\definecolor{mycolor2}{rgb}{0.85000,0.32500,0.09800}%
\definecolor{mycolor3}{rgb}{0.92900,0.69400,0.12500}%
\definecolor{mycolor4}{rgb}{0.49400,0.18400,0.55600}%
\definecolor{mycolor5}{rgb}{0.46600,0.67400,0.18800}%
\definecolor{mycolor6}{rgb}{0.30100,0.74500,0.93300}%
\begin{tikzpicture}[scale=0.5]

\begin{axis}[%
width=1.6\textwidth,
height=1.3\textwidth,
at={(0in,0in)},
scale only axis,
xmode=log,
xmin=30,
xmax=1000000,
xtick={10,100,1000,10000,100000,1000000},
xticklabels={{$\text{10}^\text{1}$},{$\text{10}^\text{2}$},{$\text{10}^\text{3}$},{$\text{10}^\text{4}$},{$\text{10}^\text{5}$},{$\text{10}^\text{6}$}},
xminorticks=true,
xlabel style={font=\color{white!15!black}, font=\large},
xlabel={$m$},
ymode=log,
ymin=1e-06,
ymax=0.01,
yminorticks=true,
ylabel style={font=\color{white!15!black}, font=\large},
ylabel={$1 - {\Pr}^{\pi^{1},\pi^{2,\circ}}( \lozenge \mathcal{S}^{R} | s_{0})$},
axis background/.style={fill=white},
axis x line*=bottom,
axis y line*=left,
xmajorgrids,
xminorgrids,
ymajorgrids,
yminorgrids,
legend style={at={(0.97,0.03)}, anchor=south east, legend cell align=left, align=left, draw=white!15!black}
]
\addplot [color=mycolor1, line width=2.0pt, mark=o, mark options={solid, mycolor1}]
  table[row sep=crcr]{%
30	2.25415562369324e-05\\
100	7.05475239435316e-05\\
300	0.000200292946528458\\
1000	0.00252784649026605\\
3000	0.00511941951335304\\
10000	0.00803418994981919\\
30000	0.00801123694332084\\
100000	0.00546942778789861\\
300000	0.00341908289748116\\
};
\addlegendentry{$n=10^5$}

\addplot [color=mycolor2, line width=2.0pt, mark=square*, mark options={solid, mycolor2}]
  table[row sep=crcr]{%
30	8.42991499616641e-06\\
100	2.6606279476149e-05\\
300	7.08733311475251e-05\\
1000	0.0002020672541172\\
3000	0.00250407956074183\\
10000	0.00692234833094585\\
30000	0.00803418956826663\\
100000	0.00801122340680793\\
300000	0.00546942449126009\\
};
\addlegendentry{$n=3\times10^5$}

\addplot [color=mycolor3, line width=2.0pt, mark=diamond*, mark options={solid, mycolor3}]
  table[row sep=crcr]{%
30	3.66644738114541e-06\\
100	8.12702868491222e-06\\
300	2.34691014394661e-05\\
1000	7.05434636734115e-05\\
3000	0.000200284682728036\\
10000	0.00250710881261307\\
30000	0.00511857080218825\\
100000	0.00803419466864008\\
300000	0.00801124427269129\\
};
\addlegendentry{$n=10^6 $}

\addplot [color=mycolor4, line width=2.0pt, mark=triangle*, mark options={solid, mycolor4}]
  table[row sep=crcr]{%
30	3.10134102621262e-06\\
100	3.78147705371035e-06\\
300	8.20560264636061e-06\\
1000	2.78557698389914e-05\\
3000	7.09695420150203e-05\\
10000	0.000200196967157451\\
30000	0.00250813254292226\\
100000	0.0069235289811771\\
300000	0.00803418744983431\\
};
\addlegendentry{$n=3\times10^6 $}

\addplot [color=mycolor5, line width=2.0pt, mark=asterisk, mark options={solid, mycolor5}]
  table[row sep=crcr]{%
30	1.83158004474215e-06\\
100	1.91138147753733e-06\\
300	3.77884294178177e-06\\
1000	8.15985798363617e-06\\
3000	2.35084725593859e-05\\
10000	7.09118862901592e-05\\
30000	0.000200269762489769\\
100000	0.00250789006280983\\
300000	0.00511645757227475\\
};
\addlegendentry{$n=10^7 $}

\addplot [color=mycolor6, dashed, line width=2.0pt]
  table[row sep=crcr]{%
30	0.003419\\
100	0.003419\\
300	0.003419\\
1000	0.003419\\
3000	0.003419\\
10000	0.003419\\
30000	0.003419\\
100000	0.003419\\
300000	0.003419\\
};
\addlegendentry{Avg. Player}

\end{axis}
\end{tikzpicture}%
		\caption{}
		\label{fig:prlose}
	\end{subfigure}
	\centering
	\caption{The value of the objective function and the probability of losing for different values of \(\sampleperstate\) and \(\samplepaths\). The dashed line in (a) marks the value of the objective function for the optimal winning policy. The dashed line in (b) marks the probability of losing under the average player's policy.}

\end{figure}

In Figure \ref{fig:klobj}, we observe that the evader is able to learn the pursuer's suboptimal policy and lower the objective function compared to the equilibrium value of the game. For lower values of \(\samplepaths/\sampleperstate\), the value of the objective function is lower than the value of the objective function under the optimal safe policy. If \(\samplepaths/\sampleperstate\) is lower, then fewer states become known and the hostile evader reaches unknown states with higher probabilities. Resultingly, the evader follows the average player's policy and incurs \(0\) payoff, which lowers the value. When \(\sampleperstate=3\times10^5\) and \(\samplepaths=10^5\), all states are unknown, and the output policy is equal to the average player's policy. In Figure \ref{fig:prlose}, if \(\samplepaths/\sampleperstate\) is low, then the probability of losing is high for the hostile evader since it follows the average player's policy with high probability. In fact, for some values of \(\samplepaths/\sampleperstate\) the probability of losing is higher than the probability that the average player loses the game. This result matches the intuition behind Lemma \ref{lemma:highreachhighcostImproved} and the \(\samplepaths/\sampleperstate\) ratio given in Proposition \ref{proposition:learning}: The learned policy may reach unknown states with higher probability compared to the average evader's policy, and to ensure that the probability of losing is low, the \(\samplepaths/\sampleperstate\) should be sufficiently high.

\section{Conclusion}
We formalized the notion of identity concealment zero-sum games and defined identity concealment games. We showed that there exists a stationary equilibrium policy pair for  identity concealment games. We then showed that a hostile player can learn a near optimal policy if the opponent is not following an equilibrium policy. In detail, we presented an algorithm that solely uses a finite number of game runs collected under the average player's policy. The output of the algorithm is a policy for the player that guarantees near optimality in the identity concealment objective and the probability of winning.

\begin{ack}            
This work was supported in part by the Air Force Research Laboratory, USA under award number FA9550-19-1-0169, the Army Research Office, USA under award number W911NF-23-1-0317, the Defense Advanced Research Projects Agency under award number D19AP00004, and the Office of Naval Research under award number N00014-23-1-2651.  
\end{ack}

\bibliographystyle{agsm}        
\bibliography{ref}           
\appendix
\section{The Proofs for Technical Results} \label{appendix:proofs}
Complete versions of the proof sketches are available at  \citep{karabag2021identity} due to the lack of space.

We use technical Lemmas \ref{lemma:divergentsubset} and \ref{lemma:lowvaluesareuseless} to prove Lemma \ref{lemma:ultimatelemma}.
\begin{lem}\label{lemma:divergentsubset}
	Let $\mathcal{D}$ be a discrete probability distribution such that $\mathcal{D}(n) \geq 0$ if $n \in \mathbb{N}$ and $\mathcal{D}(n) = 0$ otherwise, and let  $c_{1}, c_{2} \in (0, \infty)$ be arbitrary constants. Define set $D$ such that $n \in D$ if and only if $\mathcal{D}(n) > c_{1} \exp(-n c_{2})$.  If $\sum_{n = 0}^{\infty} \mathcal{D}(n) n = \infty$, we have \[\sum_{n \in D} \mathcal{D}(n) \log\left(\frac{\mathcal{D}(n)}{c_{1}\exp(-n c_{2})} \right) = \infty.\]
\end{lem}

\begin{lem} \label{lemma:lowvaluesareuseless}
	For all $n \geq 0$, the optimal value of \(\underset{x,y \in \mathbb{R}^{n}}{\min} KL(x || y)\) subject to $0 \leq x_{i} \leq y_{i}$ for all $i \in [n]$ and $\sum_{i = 1}^{n} y_{i} \leq c$, is $-c\exp(-1)$.  
\end{lem}

\begin{proof}[Proof of Lemma \ref{lemma:ultimatelemma}]
    	We partition $\mathbb{N}$ into three disjoint sets $D_{1}$, $D_{2}$, and $D_{3}$ where $n \in D_{1}$ if $\mathcal{D}^{1}(n) \leq \mathcal{D}^{2}(n) \leq c_{1}\exp(-c_{2}n)$, $n \in D_{2}$ if $\mathcal{D}^{2}(n) < \mathcal{D}^{1}(n) \leq c_{1}\exp(-c_{2}n)$, and $n \in D_{3}$ if $\mathcal{D}^{2}(n)  \leq c_{1}\exp(-c_{2}n) < \mathcal{D}^{1}(n)$. 
	
	We first lower bound the KL divergence on subsets $D_{1}$ and $D_{2}$. For subset $D_{1}$ we have 
\[		\sum_{n \in D_{1}} \mathcal{D}^{2}(n) \leq \sum_{n = 0}^{\infty} \mathcal{D}^{2}(n) \leq \sum_{n = 0}^{\infty}  \frac{c_{1}}{\exp(c_{2}n)} =  \frac{c_{1}}{\exp(c_{2}) - 1}.\]

	 By Lemma \ref{lemma:lowvaluesareuseless},  we have
	 \begin{linenomath*}
	 \begin{equation} \label{ineq:subsetc1}
\sum_{n \in D_{1}} \mathcal{D}^{1}(n) \log\left( \frac{\mathcal{D}^{1}(n)}{\mathcal{D}^{2}(n)} \right) \geq -\frac{c_{1} \exp(-1)}{\exp(c_{2}) - 1}
\end{equation}
	 \end{linenomath*}
	 since $\mathcal{D}^{1}(n) \leq \mathcal{D}^{2}(n)$ for all $n \in D_{1}$ and $\sum_{n \in D_{1}} \mathcal{D}^{2}(n) \leq  \frac{c_{1}}{\exp(c_{2}) - 1}$. For subset $D_{2}$ we have 
	 \begin{linenomath*}
	 	 \begin{equation} \label{ineq:subsetc2}
	 \sum_{n \in D_{2}} \mathcal{D}^{1}(n) \log\left( \frac{\mathcal{D}^{1}(n)}{\mathcal{D}^{2}(n)} \right) \geq 0
	 \end{equation}
	 \end{linenomath*}
 since $\mathcal{D}^{2}(n) < \mathcal{D}^{1}(n) $ and consequently $\mathcal{D}^{1}(n) \log\left( \frac{\mathcal{D}^{1}(n)}{\mathcal{D}^{2}(n)} \right) \allowbreak >0$ for all $n \in D_{2}$. 
	
	Therefore, \(KL(\mathcal{D}^{1} || \mathcal{D}^{2})\) is equal to
	\begin{linenomath*}
	\begin{subequations}
	\begin{align}
	 & \sum_{n = 0}^{\infty} \mathcal{D}^{1}(n) \log\left( \frac{\mathcal{D}^{1}(n)}{\mathcal{D}^{2}(n)} \right) \\
	&= \sum_{n \in D_{1}} \mathcal{D}^{1}(n) \log\left( \frac{\mathcal{D}^{1}(n)}{\mathcal{D}^{2}(n)} \right)+  \sum_{n \in D_{2}} \mathcal{D}^{1}(n) \log\left( \frac{\mathcal{D}^{1}(n)}{\mathcal{D}^{2}(n)} \right) \nonumber
	\\&+ \sum_{n \in D_{3}} \mathcal{D}^{1}(n) \log\left( \frac{\mathcal{D}^{1}(n)}{\mathcal{D}^{2}(n)} \right) \\
	&\geq  -\frac{c_{1} \exp(-1)}{\exp(c_{2}) - 1} + \sum_{n \in D_{3}} \mathcal{D}^{1}(n) \log\left( \frac{\mathcal{D}^{1}(n)}{\mathcal{D}^{2}(n)} \right) \label{ineq:subsetineqs}
	\\
	& \geq -\frac{c_{1} \exp(-1)}{\exp(c_{2}) - 1} + \sum_{n \in D_{3}} \mathcal{D}^{1}(n) \log\left( \frac{\mathcal{D}^{1}(n)}{c_{1} \exp(-c_{2}n)} \right) \label{ineq:defsubsetc3}
	\end{align}
\end{subequations}
	\end{linenomath*}
 where \eqref{ineq:subsetineqs} is due to \eqref{ineq:subsetc1} and \eqref{ineq:subsetc2}, and \eqref{ineq:defsubsetc3} is due to $\mathcal{D}^{2}(n) \leq c_{1}\exp(-c_{2}n)$.
	
	We note that  $n \in D_{3}$ if and only if $\mathcal{D}^{1}(n) >  c_{1}\exp(-c_{2}n)$, and  $\sum_{n = 0}^{\infty} \mathcal{D}^{1}(n) n = \infty$. By Lemma \ref{lemma:divergentsubset}, we have $\sum_{n \in D_{3}} \mathcal{D}^{1}(n) \log\left( \frac{\mathcal{D}^{1}(n)}{c_{1} \exp(-c_{2}n)} \right) = \infty$. Therefore, $KL(\mathcal{D}^{1} || \mathcal{D}^{2}) = \infty$.
\end{proof}

\begin{proof}[Proof sketch for Lemma \ref{lemma:empricalisclose}]
By Lemma 14 of \citep{strehl2009reinforcement}, with probability at least $1 - \delta_{k}/\cardstates$, we have $\| \policy^{\playertwo}(s) - \policy^{ \playertwo,*}(s) \|_{1} \leq \sqrt{\frac{2(\log(2^{\cardactions} - 2) + \log(\cardstates/\delta_{k}))}{m}}.$ Combining this with a union bound over $\states^{K}$, we get the desired result.
\end{proof}

\begin{proof}[Proof sketch for Lemma \ref{lemma:closemdpcloseoutcome}]
    We first establish that if the policy pair $(\policy^{\playerone}, \policy^{\playertwo})$ has $\left(L,\discount - \frac{\varepsilon (1-\discount)^2}{\maxcost L}\right)$-contraction, and $\| \policy^{\playertwo}(s)- \policy^{\playertwo,\circ}(s) \|_{1} \leq \frac{\varepsilon (1-\discount)^2}{\maxcost L^{2}}$ for all $\genericstate \in \states^{K}$, then $(\policy^{\playerone}, \policy^{ \playertwo,\circ})$ has $(L,\discount)$-contraction. We show this property by induction, noting that the flow difference under these two policies is at most \(\frac{\varepsilon (1-\discount)^2}{\maxcost L}\) at every \(L\)-steps. Next, we show \(\vert C(\policy^{\playerone}, \policy^{\playertwo})-C(\policy^{\playerone}, \policy^{\playertwo,\circ}) \vert  \leq {\optimalitygap}/{2}.\) Since $\| \policy^{\playertwo}(s)- \policy^{\playertwo,\circ}(s) \|_{1} \leq \frac{\varepsilon (1-\discount)^2}{\maxcost L^{2}}$, the flow difference under policy pairs \((\policy^{\playerone}, \policy^{\playertwo})\) and \((\policy^{\playerone}, \policy^{\playertwo,\circ})\) is bounded by \(\frac{\varepsilon (1-\discount)}{2\maxcost L}\). Since $(\policy^{\playerone}, \policy^{ \playertwo,\circ})$ has $(L,\discount)$-contraction, the different flow eventually reaches an end state and incurs \(0\) payoff. Due to $(L,\discount)$-contraction and the bounded payoff \(\maxcost\), this flow difference incurs at most \(\optimalitygap/2\) difference in the value functions.
\end{proof}

\begin{proof}[Proof of Lemma \ref{lemma:lowsamplelowreachChernoffImproved}]
    Let $\hat{m}^{unq}_{D}$ denote the number of sample runs that contain a transition from $D$. By Chernoff's inequality, we have 
	\begin{linenomath*}
	\begin{subequations}
	\begin{align*}
	\Pr & \left(\left|  {\Pr}^{\policy^{\average}, \policy^{ \playertwo, *}} (\lozenge D |s_{0}) - \hat{m}^{unq}_{D}/n \right| 
	\geq \sigma - \hat{m}^{unq}_{D}/n \right) \\ &\leq 2 \exp\left(-n \left(\sigma - \hat{m}^{unq}_{D}/n\right)^{2}/2\right)
	\end{align*}
\end{subequations}
	\end{linenomath*}
	where the outer probability is over the randomness of sample paths.

	Note that $\hat{m}^{unq}_{D} \leq \hat{m}_{D} \leq m'$ since \(\hat{m}^{unq}_{D}\) is the number of paths with a transition from \(D\) and \(\hat{m}_{D}\) is the total number of transitions from \(D\). Therefore, we have
	\begin{linenomath*}
	\begin{subequations}
	\begin{align*}
	\Pr& \left( {\Pr}^{\policy^{\average}, \policy^{ \playertwo,*}} (\lozenge D |s_{0})  \geq \sigma \right) 	
	\\
	&\leq \Pr\left( \left| {\Pr}^{\policy^{\average}, \policy^{ \playertwo,*}} (\lozenge D |s_{0}) - \hat{m}^{unq}_{D} /n \right|  \geq \sigma - \hat{m}^{unq}_{D}/n \right)
	\\
	&\leq 2\exp\left(-n \left( \sigma - \hat{m}^{unq}_{D}/n \right)^{2}/2\right)
	\\
	& \leq 2  \exp\left(-n \left( \sigma - m'/n\right)^{2}/2\right)
	\end{align*}
\end{subequations}
	\end{linenomath*}
	which yields the desired result.
\end{proof}

\begin{proof}[Proof of Lemma \ref{lemma:highreachhighcostImproved}]
    
    Let \(\rho^{1} = {\Pr}^{\policy^{\playerone}, \policy^{ \playertwo,\circ}} (\lozenge D |s_{0}) \) and \(\rho^{\average} = {\Pr}^{\policy^{\average}, \policy^{ \playertwo,\circ}} (\lozenge D |s_{0})\).
	Note that \( C(\policy^{1}, \policy^{2,\circ}) = 
		KL\left(\policy^{\playerone}, \policy^{\playertwo, \circ} || \policy^{\average}, \policy^{\playertwo,\circ}\right) \allowbreak\geq KL\left(Ber\left(\rho^{1} \right) || Ber\left(\rho^{\average} \right) \right) \)
 due to the data processing inequality. Therefore, it suffices to show that if  $\rho^{1}  > \rho^{\average}$ and $\rho^{1}  > \frac{ v^{*} + \log(2) + \varepsilon}{-\log\left(\rho^{\average} \right)},$ then $KL\left( Ber\left(\rho^{1} \right) || Ber\left(\rho^{\average} \right) \right) > v^{*} + \varepsilon.$
	
	We have 
	\begin{linenomath*}
	\begin{subequations}
	\begin{align}
	&KL\left(Ber\left(\rho^{\playerone} \right) || Ber\left(\rho^{\average} \right) \right)
	\\ &= \rho^{\playerone}  \log\left( \frac{\rho^{\playerone} }{\rho^{\average} } \right) + \left(1 - \rho^{\playerone} \right) \log\left( \frac{1 - \rho^{\playerone}}{1 - \rho^{\average}} \right)
	\\ &\geq \rho^{\playerone} \log\left( \frac{\rho^{\playerone}}{\rho^{\average}} \right) + \left(1 - \rho^{\playerone} \right) \log\left( 1 - \rho^{\playerone} \right)
	\\ &\geq - \rho^{\playerone} \log\left( \rho^{\average} \right) -\log(2) \label{ineq:xlogxlowerbounded}
	\end{align}
\end{subequations}
	\end{linenomath*}
	where  \eqref{ineq:xlogxlowerbounded} is because of that $\min x \log(x) + (1-x) \log(1-x)$ subject to $x \in [0,1]$ is $-\log(2)$. We get the desired result by rearranging the terms in \eqref{ineq:xlogxlowerbounded}.
\end{proof} 

\begin{proof}[Proof of Lemma \ref{lemma:divergentsubset}]
    
	Fix arbitrary constants $c_{1}, c_{2} \in (0, \infty)$. We partition $\mathbb{N}$ into three disjoint subsets $D_{1}$, $D_{2}$, and $D_{3}$ such that $n \in D_{1}$ if and only if $\mathcal{D}(n) \leq c_{1}\exp(-nc_{2})$, $n \in D_{2}$ if and only if $ c_{1}\exp(-nc_{2}) \leq \mathcal{D}(n) < c_{1}\exp(-nc_{2}/2)$, and $n \in D_{3}$ otherwise. Also define $D := D_{2} \cup D_{3}.$
	
	\(\sum_{n \in D_{1} \cup D_{2}} \mathcal{D}(n) n \allowdisplaybreaks \leq \sum_{n \in D_{1} \cup D_{2}} c_{1} \exp(-n c_{2}/2) n \allowbreak \leq \allowdisplaybreaks \sum_{n = 0}^{\infty} \allowbreak c_{1}\exp(-nc_{2}/2) n \allowbreak = \frac{c_{1}\exp(c_{2}/2)}{(\exp(c_{2}/2) - 1)^{2}} < \infty\) since $c_{1}, c_{2} \in (0, \infty)$.
	
	Since $\mathcal{D}(n) n \geq 0$ for all $n \in \mathbb{N}$, we have \[\sum_{n=0}^{\infty} \mathcal{D}(n) n = \sum_{n \in  D_{1} \cup D_{2}} \mathcal{D}(n) n + \sum_{n \in D_{3}} \mathcal{D}(n) n.\] Since $\sum_{n=0}^{\infty} \mathcal{D}(n) n$ diverges and $\sum_{n \in D_{1} \cup D_{2}} \mathcal{D}(n) n$ converges, we must have $\sum_{n \in D_{3}} \mathcal{D}(n) n = \infty$.  
	
	We have 
	\begin{linenomath*}
	\begin{subequations}
		\begin{align}
			\sum_{n \in D}& \mathcal{D}(n) \log\left(\frac{\mathcal{D}(n)}{c_{1}\exp(-n c_{2})} \right) 
			\\
			&=	\sum_{n \in D_{2} \cup D_{3}} \mathcal{D}(n) \log\left(\frac{\mathcal{D}(n)}{c_{1}\exp(-n c_{2})} \right)
			\\
			&\geq \sum_{n \in D_{3} } \mathcal{D}(n) \log\left(\frac{\mathcal{D}(n)}{c_{1}\exp(-n c_{2})} \right) \label{ineq:defc3}
			\\
			&\geq \sum_{n \in D_{3} } \mathcal{D}(n) \log\left(\frac{c_{1}\exp(-n c_{2}/2)}{c_{1}\exp(-n c_{2})} \right) \label{ineq:defc32} 
			\\
			&=\infty \label{ineq:divergent}
		\end{align}
	\end{subequations}
\end{linenomath*}
where \eqref{ineq:defc3} is due to $\mathcal{D}(n) \geq c_{1}\exp(-nc_{2})$ for all $n \in D_{2}$, \eqref{ineq:defc32} is due to $\geq c_{1}\exp(-nc_{2}/2)$ for all $n \in D_{3}$, and \eqref{ineq:divergent} is due to $\sum_{n \in D_{3}} \mathcal{D}(n) n = \infty$ and $c_{2} > 0$.
\end{proof}

\begin{proof}[Proof sketch for Lemma \ref{lemma:lowvaluesareuseless}]
     The proof follows from the convexity of KL divergence and the first-order optimality condition.
\end{proof}

\end{document}